\documentclass[preprint,showpacs,preprintnumbers]{revtex4}
\usepackage{amssymb}
\usepackage{amsmath}
\usepackage{amsfonts}
\usepackage{graphicx}
\newcommand\T{\rule{0pt}{2.6ex}}
\newcommand\B{\rule[-1.2ex]{0pt}{0pt}}
\usepackage{dcolumn}
\usepackage{bm}
\usepackage{appendix}

\begin{document}
\preprint{}

\title{Theoretical coarse-graining approach to bridge length scales
in diblock copolymer liquids}
\author{E. J. Sambriski and M. G. Guenza}
\affiliation{Department of Chemistry and Institute of Theoretical
Science, University of Oregon, Eugene, Oregon 97403, USA}
\date{\today}

\begin{abstract}
\noindent A microscopic theory for coarse graining diblock copolymers
into dumbbells of interacting soft colloidal particles has been
developed, based on the solution of liquid-state integral equations.
The Ornstein-Zernike equation is solved to provide a mesoscopic
description of the diblock copolymer system at the level of block
centers of mass, and at the level of polymer centers of mass.
Analytical forms of the total correlation functions for block-block,
block-monomer, and center-of-mass pairs are obtained for a liquid of
structurally symmetric diblock copolymers as a function of temperature, density, chain length, and chain 
composition. The theory correctly predicts 
thermodynamically-driven segregation of diblocks into microdomains as a function
of temperature ($chi$ parameter). The coarse-grained description contains contributions
from density and concentration fluctuations, with the latter becoming
dominant as temperature decreases.  Numerical
calculations for the block coarse-grained total correlation functions,
as a function of the proximity of the system to its phase transition,
are presented.  Comparison with united atom molecular dynamics
simulations are carried out in the athermal regime, where simulations
and theory quantitatively agree with no need of adjustable parameters.
\end{abstract}


\maketitle

\section{Introduction}
\label{SX:INTR}
\noindent One of the challenges in understanding the properties of
polymeric materials stems from the necessity of developing theoretical
approaches that can describe in a comprehensive manner properties
observed at many different length scales.  The presence of several
length scales in which relevant phenomena take place leads to the
complex nature of the liquid, rendering its treatment a difficult
matter \cite{complexfluids,complexfluids1,complexfluids2}.  Already in
the description of the structure of homopolymer melts, two length
scales need to be considered, which correspond to the monomer
statistical segment length, $\sigma$, and the overall polymer
dimension, i.e.\ its radius of gyration, $R_g= \sigma (N/6)^{1/2}$,
where $N$ is the total number of monomers in the chain.  For diblock
copolymers, the theoretical treatment is further complicated by the
presence of a new length scale, which is intramolecular in character
and corresponds to the size of a block.  Diblock copolymers are
macromolecules in which a homopolymer chain of $N_A=f N$ monomers of
type $A$ is chemically bound to a second homopolymer chain of
different chemical structure containing $N_B=(1-f)N$ monomers of type
$B$, with $N=N_A+N_B$.  The block size is defined by its respective
radius of gyration such that, for example, the spatial dimension of
the block composed of $A$-type monomers is given by $R_{gA}=\sigma
(N_A/6)^{1/2}$.

Diblock copolymers are systems of great interest for their
technological applications \cite{diblock,balsara}.  Since the two
blocks are chemically different, these experience a repulsive
interaction that would encourage phase separation at low temperatures
where entropy cannot balance enthalpic effects.  However, the chemical
bond existent between the two blocks prevents a complete separation of
the two phases.  As a consequence, at low temperatures block copolymer
liquids undergo a microphase transition from disordered systems to
ordered microstructures of nanoscopic size, namely the microphase
separation transition (MST). The length scale characterizing the size
of the microphase is of the order of the block radius of gyration.

With the purpose of developing the technology to produce micro-ordered
structures of well-controlled size and shape, an understanding is
required of the processes that drive the formation of micro-ordered
phases under different thermodynamic conditions of temperature $T$ and
density $\rho$, as well as different chain composition $f$, monomer
structure $\sigma$, and degree of polymerization $N$.

Computer simulations serve as an extremely powerful tool to
investigate phenomena in complex fluids \cite{simul}.  However, the
limited power of present-day hardware does not allow for the
simultaneous study of all length scales of interest.  One way to
overcome this problem is through ``multiscale modeling,'' where a set
of simulations is performed at different levels of coarse-graining of
the original system and in a subsequent step, information from
different length scales is combined to provide the complete physical
picture \cite{simul1}.  With such an approach, however, the challenge
is not only to find the appropriate computational technique for each
length scale simulated, but also to know (\textit{i}) the proper
effective potential acting between coarse-grained units needed to
carry out the simulations \cite{DAUTN,KRMER,Maass,BRIELS}, and (\textit{ii})
the proper procedure for combining information from different scales
of modeling once simulation data is acquired.

First-principles theoretical models apt to coarse grain diblock
copolymer liquids at different length scales of interest provide the
potentials needed as an input in multiscale simulations, as well as
the formal framework to combine information obtained from simulations
of the liquid coarse grained at different length scales
\cite{K2002,K2003,K2004,HNCOL,MNBDY}.  In a series of recent papers,
we developed a coarse-graining approach that maps liquids of
homopolymer chains into liquids of soft colloidal particles, providing
a formal analytical ``transcale'' procedure \cite{YAPRL,MLTEX,JOPCM}.
Each effective soft colloid is centered on the polymer center of mass
and interacts through a Gaussian repulsive potential, which results in
our formalism from the solution of Ornstein-Zernike integral
equations.  We later extended the same approach to describe the
coarse-graining of homopolymer mixtures \cite{YAPRL,BLNDS}.  Computer
simulations, performed by us, of coarse-grained polymer liquids and
mixtures, where molecules interact by means of the derived effective
pair potential, have been shown to reproduce quantitatively the
structure and dynamics of the liquid at the center-of-mass level,
while requiring considerably shorter computational time than that
needed to perform united atom simulations \cite{YAPRL,MLTEX,JOPCM,BLNDS}.

In the present work, our approach is further developed to address the
problem of modeling a melt of diblock copolymers as a liquid of
interacting soft colloidal dumbbells.  Each dumbbell represents one
macromolecule composed of two effective soft colloidal particles,
which in turn are sized according to the radius of gyration of each
block and centered on center-of-mass coordinates of each block.  Three
different length scales are formally related, which correspond to
coarse-graining the molecule at the monomer (the statistical segment
length, $\sigma$), block (the radius of gyration of block $A$,
$R_{gA}$), and polymer (the polymer radius of gyration, $R_g$) scales.
In this way, our theory represents a minimal {\it intra}molecular
mesoscopic model of polymeric liquid structures.

There has been a growing interest in providing models for coarse-graining 
block copolymers chains.\cite{Andelman,muller,Liang} For example, building blocks of supermolecular structures, 
such as cellular membranes, could be modeled as self-assembling
block copolymers chains.\cite{Pier}
A recent paper proposes a model of coarse-graining for a symmetric diblock  copolymer
similar to ours, as the chain is modeled as two soft blobs, tethered by an entropic spring.\cite{Addison} 
The  blobs have equal size, and the coarse-grained total distribution functions are 
calculated numerically from a Monte Carlo
simulation of diblock copolymers described at the monomer level. Monomers occupy the sites of a simple cubic lattice, with bond along the x-, y-, or z-directions. The two blocks individually
are modeled as if they where in theta solvent, while the interaction between them 
is self avoiding. The numerical inversion procedure 
to derive the coarse-grained potential is performed in the athermal regime. 
As the authors point out in the paper, their model is "highly simplified", which
proves the difficulty in treating intramolecular coarse-graining.
The model, coupled with  a reference interaction site model (RISM) 
and a random-phase approximation closure,
predicts the mean-field clustering of diblock copolymers in a selective solvent.\cite{Hansen}

In this paper, we provide an analytical solution for the coarse-grained total distribution functions for a liquid of diblock copolymers represented as dumbbells of soft colloidal particles. Our model differs from the one presented in Ref.(\cite{Addison}) in several ways. In our case, the size of the two "blobs" varies 
depending on the chain composition, $f$, degree
of polymerization, $N$, and segment length, $\sigma$. Moreover, repulsive interactions between segments of different chemical nature are quantified by the interaction parameter, $\chi_{\mathit{eff}} $. 
Concentration-fluctuation stabilization enters through the polymer reference interaction site model (PRISM) theory for the monomer-level description,\cite{PRISM,eddavid,eddavid1}
and deviations from mean-field theory\cite{LEIBL} are predicted by the coarse-grained approach as well.
The two blocks follow Gaussian intramolecular statistics, which is a good approximation for copolymer melts, when each block has a degree of polymerization $N_{\alpha} > 30$, with $\alpha \in {A,B}$, and for the region in the phase diagram from the high-temperature to the weak segregation regime ($\chi_{eff}N << 10.5$ for symmetric composition $f=0.5$), where the system is isotropic. Numerical mean field theory studies suggest coil stretching is not significant even below the order disorder transition until a strong segregation regime is entered, where $\chi_{eff}N \ge 100$.\cite{44a,44b}
Analytical intermolecular total correlation functions between like and unlike coarse-grained blocks
are predicted  by our formalism as a function of chain composition, block size, density, temperature, as well as density- and concentration-fluctuation screening lengths.

One advantage of our approach is that it is analytical.  Since a
coarse-grained description is obtained by performing statistical
averages over local (small-scale) degrees of freedom, it translates
the energy of the system into the free energy of the renormalized
fluid.\cite{Likos}  In this way, the obtained total correlation functions  and related
effective potentials, are  functions of all
characteristic physical parameters defining the system under
consideration, such as temperature $T$, total site density $\rho$, and
degree of polymerization $N$.  For a diblock copolymer, the relevant
parameters also include chain composition, $f$, and the interaction
parameter $N\chi_{\mathit{eff}} \propto T^{-1}$.  The parameter
$\chi_{\mathit{eff}}$ defines the proximity of the system to its
order-disorder transition.  For each set of parameters, total correlation functions and
free energy change.  As a result, a numerical solution of the
coarse-grained description obtained from microscopic simulations requires performing
a number of simulations equal to the number of combinations of those
parameters, partially defeating
the computational gains of a coarse-grained description.

The material in this paper is organized in the following manner.  We
start in Section \ref{SX:BLOK} with a derivation from the
Ornstein-Zernike equation of the total correlation functions of
diblock copolymer liquids coarse-grained at the block level.  Section
\ref{SX:THEO} provides a theoretical description of diblock copolymer
melts at the monomer level.  In Section \ref{SX:SYMM}, we present an
analytically tractable solution in reciprocal space for a structurally
symmetric diblock copolymer melt.  The Fourier transform of the
resulting expressions leads to analytical solutions of intermolecular
total block-monomer and block-block pair correlation functions in real
space, treated in Section \ref{SX:SYMMR}.  Our analytical approach for
coarse-graining the diblock copolymer liquid at the center-of-mass
level and comparisons with the corresponding presentation of the
homopolymer melt are discussed in Section \ref{SX:DCOM}.  Finally, our
theoretical development in the {\it athermal} limit is compared with
united atom molecular dynamics simulations of a polyethylene melt in
Section \ref{SX:COMP}, while temperature-dependent model calculations
are presented in Section \ref{SX:MODL}.  The paper concludes with a
brief discussion and Appendices, where the auxiliary functions entering the 
exact solution of the
total correlation functions, as well as the
treatment of the block coarse-graining of a compositionally {\it
symmetric} diblock, are presented.

\section{An integral equation approach to coarse-grain diblock
copolymers at the block length scale}
\label{SX:BLOK}
\noindent In this section, we derive the general expressions for the
total pair correlation functions of the diblock copolymer liquid,
coarse-grained at the level of the block length scale.  The formalism
is completely general and applies to any diblock copolymer system.
The structure of a diblock copolymer liquid is characterized well by
static correlation functions, which sample fluctuations of monomer
units in the fluid.  Given the position of monomer $a$ belonging to
block $\alpha\in\{A,B\}$ comprising a chain $j$ as
$\vec{r}\vphantom{r}^{\,j}_{a}$, monomer fluctuations at a specific
wave vector $\vec{k}$ are represented by $\rho^{j}_{a}(\vec{k}) =
e^{i\vec{k}\cdot\vec{\vphantom{k} r}\vphantom{r}^{\,j}_{a}}$.  Since
all chains are assumed to be equivalent, we henceforth discard the
chain index $j$.  Density fluctuations for a generic monomer inside a
block $\alpha$ are defined as $\rho_{\alpha}(\vec{k}) =
\sum^{N_\alpha}_{a = 1} e^{i\vec{k}\cdot\vec{\vphantom{k}r}_{a}}$,
while $\rho_{\alpha}(\vec{k})^*$ defines its complex conjugate. The
two-point density correlation functions, which describe the liquid
structure, are given by the partial structure factor
\begin{equation}
S^{mm}_{\alpha\beta}(k) =
\frac{1}{N}\langle\rho_\alpha(\vec{k}) \cdot
\rho_\beta(\vec{k})^*\rangle =
\mathit{\Omega}^{mm}_{\alpha\beta}(k) +
H^{mm}_{\alpha\beta}(k) \, ,
\end{equation}

\noindent which includes static correlations between monomers
belonging to the same chain, $\mathit{\Omega}^{mm}_{\alpha\beta}(k)$,
i.e.\ the intramolecular static structure factor, and correlations
between monomers belonging to different chains,
$H^{mm}_{\alpha\beta}(k)$, i.e.\ the intermolecular structure factor.
Since the liquid is spatially homogeneous and isotropic, the structure
factors depend only on the modulus of the wave vector,
$|\vec{k}|\equiv k$.

In our coarse-grained description, each block comprising the chain is
mapped onto an effective particle, centered on the position of the
block center-of-mass coordinate, $\vec{R}_{b\alpha} =
N_\alpha^{-1}\sum^{N_\alpha}_{a = 1}\vec{r}_a$.  Fluctuations from the
center of mass of block $\alpha$ are defined as
$\rho_{b\alpha}(\vec{k}) = e^{i\vec{k}\cdot\vec{R}_{b\alpha}}$, and
the partial structure factor becomes
\begin{equation}
S^{bb}_{\alpha\beta}(k) =
\frac{1}{2}\langle\rho_{b\alpha}(\vec{k})\cdot
\rho_{b\beta}(\vec{k})^*\rangle =
\mathit{\Omega}^{bb}_{\alpha\beta}(k) + H^{bb}_{\alpha\beta}(k) \, ,
\end{equation}

\noindent which requires knowledge of both intra- and intermolecular
correlations.  Block structure factors are normalized by the number of
blocks in the chain, which is two for a diblock copolymer molecule.

By analogy, block-monomer two-point correlation functions are given by
\begin{equation}
S^{bm}_{\alpha\beta}(k)
= \frac{1}{2}\langle\rho_{b\alpha}(\vec{k})\cdot
\rho_{\beta}(\vec{k})^*\rangle
= \mathit{\Omega}^{bm}_{\alpha\beta}(k) + H^{bm}_{\alpha\beta}(k) \, ,
\end{equation}

\noindent which for compositionally asymmetric chains,
$\mathit{\Omega}^{bm}_{\alpha\beta}(k) \neq
\mathit{\Omega}^{bm}_{\beta\alpha}(k)$, $H^{bm}_{\alpha\beta}(k) \neq
H^{bm}_{\beta\alpha}(k)$, and consequently $S^{bm}_{\alpha\beta}(k)
\neq S^{bm}_{\beta\alpha}(k)$.

Intra- and intermolecular structure factors are related through the
Ornstein-Zernike equation, which has the general matrix formula
\begin{equation}
\mathbf{H}(k) = \mathbf{\Omega}(k)\mathbf{C}(k)\mathbf{S}(k) \, .
\label{EQ:OZEQ}
\end{equation}

\noindent Here, $\mathbf{C}(k) = \mathbf{\Omega}^{-1}(k) -
\mathbf{S}^{-1}(k)$ is the intermolecular direct pair correlation
function matrix.  In our description, the generalized Ornstein-Zernike
equation includes contributions from ``real'' monomeric sites ($m$)
and ``auxiliary'' sites positioned on the block center-of-mass site
($b$). 

At the block level description, matrices share similar arrangements. As an example we show the partial static
structure factor, $\mathbf{S}(k) =
\mathbf{\Omega}(k) + \mathbf{H}(k)$, which is defined as
\begin{equation}
\mathbf{S}(k) =
\left[
\begin{array}{c|c}
\mathbf{S}^{mm} & \mathbf{S}^{bm} \\
\hline
\left(\mathbf{S}^{bm}\right)^{T^{\vphantom{T}}}
& \mathbf{S}^{bb}
\end{array}
\right] =
\left[
\begin{array}{cc|cc}
S^{mm}_{AA} & S^{mm}_{AB} & S^{bm}_{AA} & S^{bm}_{AB} \\
S^{mm}_{BA}
& S^{mm}_{BB} & S^{bm}_{BA} & S^{bm}_{BB} \\
\hline
S^{bm^{\vphantom{T}}}_{AA}
& S^{bm}_{BA} & S^{bb}_{AA} & S^{bb}_{AB} \\
S^{bm}_{AB} & S^{bm}_{BB} & S^{bb}_{BA} & S^{bb}_{BB}
\end{array}
\right] \, ,
\end{equation}

\noindent where we omit the variable $k$ to simplify the notation.
In an analogous way we define the intramolecular structure factor matrix, which contains the
correlation between real sites, $ \mathbf{\Omega}^{mm}_{\alpha \beta}=\rho \omega^{mm}_{\alpha \beta}$, auxiliary sites, $\mathbf{\Omega}^{bb}_{\alpha \beta}=\rho_b \omega^{bb}_{\alpha \beta}$,
and the corresponding cross contributions, $\mathbf{\Omega}^{bm}_{\alpha \beta}=\rho_b \omega^{bm}_{\alpha \beta}$.
Here, the block number density is $\rho_b = 2\rho/N$, and $\omega^{bb}_{AA}(k) = \omega^{bb}_{BB}(k) = 1/2$ for a diblock copolymer.  Furthermore, for blocks of different type, $\alpha \neq
\beta$, we have that $\omega^{mm}_{\alpha\beta}(k) =
\omega^{mm}_{\beta\alpha}(k)$ and $\omega^{bb}_{\alpha\beta}(k) =
\omega^{bb}_{\beta\alpha}(k)$, while $\omega^{bm}_{\alpha\beta}(k)
\neq \omega^{bm}_{\beta\alpha}(k)$.

Consistent with its intramolecular counterpart, the matrix of total
intermolecular pair correlation functions contains the correlation term for the real sites,
$\mathbf{H}^{mm}_{\alpha \beta}=\rho_\alpha \rho_\beta h^{mm}_{\alpha \beta}$, auxiliary sites
$ \mathbf{H}^{bb}_{\alpha \beta}=\rho_{b \alpha} \rho_{b \beta} h^{bb}_{\alpha \beta}$, and cross contributions $\mathbf{H}^{bm}_{\alpha \beta} = \rho_{b \alpha} \rho_{b \beta} h^{bm}_{\alpha \beta}$.

\noindent Here the number density of monomers $A$ is $\rho_A = f\rho$
and $\rho_B = (1-f)\rho$, while the number density of blocks of type
$A$ (or $B$) is $\rho_{bA} = \rho_{bB} = \rho_b/2$ for a diblock
copolymer.

Finally, the intermolecular direct pair correlation function matrix includes the usual assumption that auxiliary sites are
not directly correlated with either real or other auxiliary sites.
In this theoretical framework, we obtain the correlation in fluctuations of the intermolecular
block-monomer function
\begin{align}
\mathbf{H}^{bm}(k) & =
\mathbf{\Omega}^{bm}(k)
\left[\mathbf{\Omega}^{mm}(k)\right]^{-1}
\mathbf{H}^{mm}(k) \, ,
\label{EQ:KBMLK}
\end{align}

\noindent and the intermolecular block-block function
\begin{align}
\mathbf{H}^{bb}(k) & =
\mathbf{\Omega}^{bm}(k)
\left[\mathbf{\Omega}^{mm}(k)\right]^{-1}
\mathbf{H}^{mb}(k)\, ,
\label{EQ:KBLK}
\end{align}

\noindent where $\mathbf{H}^{bm}(k) =
\left[\mathbf{H}^{mb}(k)\right]^T$, since $h^{bm}_{\alpha\beta}(k)
\neq h^{bm}_{\beta\alpha}(k)$ whenever $\alpha \neq \beta$ with
$\alpha,\beta \in \{A,B\}$.  The general block-block relation reads

%\begin{equation}
%\mathbf{H}^{bb}(k) =
%\mathbf{\Omega}^{bm}(k)
%\left[\mathbf{\Omega}^{mm}(k)\right]^{-1}
%\left[\mathbf{H}^{mm}(k)\right]^T
%\{\left[\mathbf{\Omega}^{mm}(k)\right]^{-1}\}^T
%\left[\mathbf{\Omega}^{bm}(k)\right]^T \, ,
%\end{equation}

\begin{equation}
\mathbf{H}^{bb}(k) =
\mathbf{\Omega}^{bm}(k)\left[\mathbf{\Omega}^{mm}(k)\right]^{-1}
\mathbf{H}^{mm}(k)\left[\mathbf{\Omega}^{mm}(k)\right]^{-1}
\mathbf{\Omega}^{mb}(k)\, ,
\label{EQ:BBLK}
\end{equation}
where we used the property that 
both $\mathbf{H}^{mm}(k)$ and $\mathbf{\Omega}^{mm}(k)$ are symmetric
matrices, along with the definition $\mathbf{\Omega}^{mb}(k) =
\left[\mathbf{\Omega}^{bm}(k)\right]^T$.
Eq.(\ref{EQ:KBLK}) formally relates the total correlation function for a coarse-grained diblock copolymer, represented as a dumbbell of two soft colloidal particles, to the total correlation function and intramolecular structure factor of the monomer-level description.

Eqs.\ (\ref{EQ:KBMLK}) and (\ref{EQ:BBLK}) are further simplified using the fact that $h^{bm}_{AB}(k) \neq h^{bm}_{BA}(k)$
and $h^{bb}_{AB}(k)= h^{bb}_{BA}(k)$.  The 
block-monomer total correlation functions follows the general expression
\begin{eqnarray}
h^{bm}_{\alpha \beta}(k)  =  \frac{2}{\omega(k)} 
\Bigg\{& - & (1-f) \Big[ \omega^{mm}_{AB} \omega^{bm}_{\alpha A} h^{mm}_{B \beta}
 -  \omega^{mm}_{AA} \omega^{bm}_{\alpha B} h^{mm}_{B\beta} \Big]  \nonumber \\
& + & f \Big[ \omega^{mm}_{BB} \omega^{bm}_{\alpha A} h^{mm}_{A \beta} - 
\omega^{mm}_{AB} \omega^{bm}_{\alpha B} h^{mm}_{A \beta} \Big]
\Bigg\} 
\label{EQ:HBMS1}
\end{eqnarray}

\noindent where 
\begin{equation}
\omega(k) = \omega^{mm}_{AA}(k)\omega^{mm}_{BB}(k) -
\left[\omega^{mm}_{AB}(k)\right]^2 \, ,
\label{EQ:omega}
\end{equation}

\noindent and $\alpha \beta \in{AB}$, while  the
block-block total correlation functions follows
\begin{eqnarray}
h^{bb}_{\alpha \beta}(k) & = & (1-f)^2 h^{mm}_{BB} A_{\alpha \beta}^{(1)} -f (1-f) h^{mm}_{AB} A_{\alpha \beta}^{(2)} + f^2 h_{aa}^{mm} A_{\alpha \beta}^{(3)}\, ,
\label{EQ:HBMS}
\end{eqnarray}
with
\begin{eqnarray}
A_{\alpha \beta}^{(1)} & = & (\omega_{AB}^{mm})^2 \omega_{\alpha A}^{bm} \omega_{\beta A}^{bm} -
\omega^{mm}_{AA} \omega^{mm}_{AB} ( \omega ^{bm}_{\alpha b} \omega^{bm}_{\beta A} +
\omega^{bm}_{\alpha A} \omega^{bm}_{\beta B}) +(\omega_{AA}^{mm})^2 \omega^{bm}_{\alpha B} \omega^{bm}_{\beta B} \, , \nonumber \\
A_{\alpha \beta}^{(2)} & = &   2 \omega^{mm}_{AB} \omega^{mm}_{BB} \omega^{bm}_{\alpha A} \omega^{bm}_{\beta A}
- [\omega^{mm}_{AA} \omega^{mm}_{BB} + (\omega^{mm}_{AB})^2][\omega^{bm}_{\alpha B}\omega^{bm}_{\beta A}+ \omega^{bm}_{\alpha A} \omega ^{bm}_{\beta B}] + 2 \omega ^{mm}_{AA} \omega^{mm}_{AB} \omega^{bm}_{\alpha B} \omega^{bm}_{\beta b} \, , \nonumber \\
A_{\alpha \beta}^{(3)} & = &  (\omega^{mm}_{BB})^2 \omega^{bm}_{\alpha A} \omega^{bm}_{\beta A} - \omega^{mm}_{AB} \omega^{mm}_{BB}
[\omega^{bm}_{\alpha B} \omega^{bm}_{\beta A} + \omega^{bm}_{\alpha A} \omega^{bm}_{\beta B}] + (\omega^{mm}_{AB})^2 \omega^{bm}_{\alpha B} \omega^{bm}_{\beta B}\ .
\end{eqnarray}

Since no particular structure of the diblock copolymer has been
assumed thus far, Eqs.\ (\ref{EQ:HBMS1}) and (\ref{EQ:HBMS}) are
completely general and hold for any model of diblock copolymer chains,
including diblock copolymers with asymmetric chain segments, as well
as any general form of an interaction potential.  From the knowledge
of the pair correlation functions, obtained from the Fourier transform
of $\mathbf{\Omega}(k)$ and $\mathbf{H}(k)$, all thermodynamic
properties of a polymer liquid can be formally derived
\cite{hansenmacd}.

In the following sections, we present an {\it analytical} solution for
the intermolecular block-monomer and block-block correlation functions
for a diblock copolymer liquid.  We assume a structurally and
interaction symmetric diblock with variable chain composition. The
molecule is modeled  as a Gaussian ``thread'' of infinite length and
vanishing thickness. This model allows for an analytical solution of
the total correlation functions in real and reciprocal spaces, as  a
function of the thermodynamic parameters of the system.

\section{Monomer level description of diblock copolymer liquids}
\label{SX:THEO}
\noindent The coarse-graining formalism presented in Section
\ref{SX:BLOK} is simplified when structurally and interaction
symmetric diblock copolymers are investigated. For these model
systems, segments of different chemical nature are assumed to have
equivalent statistical lengths, $\sigma_A = \sigma_B = \sigma$, while
the specific chemical nature of the block enters as an effective
persistence length, through the block radius of gyration.  Segments of
like species interact through the potentials $v_{AA} \approx v_{BB}$,
while unlike species repel each other through $v_{AB}$.  At high
temperatures, entropic effects dominate over enthalpic contributions,
and block copolymer liquids resemble closely liquids of homopolymer
molecules.  As the temperature decreases, the effective repulsive
potential $ \chi_{\mathit{eff}} = v_{AA} + v_{BB} - 2v_{AB}$ increases
as $N \chi_{\mathit{eff}} \propto T^{-1}$, leading to the phase
separation transition. This phase transition is characterized  by a
dramatic increase of the collective concentration fluctuation static
structure factor, $S^{\phi}(k^*) $, at a specific length scale, $k^*$.
At the temperature of the phase transition, only certain fluctuations
become anomalously large and the liquid segregates on a molecular
length scale on the order of the overall size of the molecule, $k^*
\sim R_g^{-1}$.  This remarkable property of copolymer liquids is due
to the fact that, because of the connectivity between different
blocks, even complete segregation cannot lead to macroscopic phase
separation, as occurs in mixtures of two chemically different
homopolymer melts. Since even at high temperatures, $S^{\phi}(k)$
presents a peak due to the finite molecular size of the block
copolymer chain, the peak position is largely independent of
temperature.

The first theoretical approach to describe the microphase separation
transition was a mean-field theory developed by  Leibler \cite{LEIBL}.
The theory is built on the expansion of the free-energy density of an
ordered phase in powers of the order parameter, defined as the average
deviation from the uniform distribution of monomers.  The theory
predicts a second-order phase transition for the symmetric mixture ($f
= 0.5$) and a first-order transition for asymmetric systems.
Mean-field approaches are usually less accurate in the vicinity of the
transition, where fluctuation corrections to the mean-field theory can
change drastically the phase diagram.  In the case of diblock
copolymer melts, these corrections modify the predicted second-order
phase transition for the symmetric diblock into a first-order phase
transition.

A fluctuation corrected mean-field approach was later derived by
Fredrickson and Helfand \cite{FRDIK} by implementing Brazovskii's
Hartree approximation of Landau-Ginzburg field theory to treat diblock
copolymer systems \cite{BRAZO}.  The fluctuation corrected approach
recovers Leibler's results in the limit of infinite chain length.
Fredrickson-Helfand predictions have been found to agree well with
scattering experiments in the whole range of temperatures encompassing
disordered to weakly ordered phases across the microphase
separation.  Since these approaches focus on the single-chain free
energy and the liquid is assumed to be incompressible, fluctuation
stabilization is mainly of entropic origin.

Schweizer and coworkers developed an integral equation description of
block copolymer melts that formally recovers the scaling behaviors
obtained in field theories with only small differences
\cite{eddavid,eddavid1,MDCMP,MDDYN,MDDYN1}.  In this case, however, the
stabilization of the disordered state close to the MST is of enthalpic
origin.  Moreover, since the theory does not enforce incompressibility
as a starting point in the treatment, the resulting structure factor
contains contributions from both density and concentration
fluctuations.  This is consistent with the physical picture that in
block copolymer melts far from their MST, the concentration
fluctuation contribution is negligible while density fluctuations can
still occur.

In this work, we adopt an integral equation approach to describe the
block copolymer structure at the monomer level, extending Schweizer's
theory.  This liquid-state description is largely compatible with the
fluctuation-corrected mean-field approach, and has the advantage of
providing a theoretical framework that is formally consistent with the
procedure presented in the previous section. In this way, the approach
presented here builds on, and merges two well-developed theoretical
fields involving (\textit{i}) the extension of integral equation
approaches to treat complex molecular liquids
\cite{blockschweizer,PRISM}, and (\textit{ii}) procedures to coarse
grain macromolecular liquids
\cite{YAPRL,MLTEX,JOPCM,K2002,HNCOL,MNBDY}.

We focus on a structurally and interaction symmetric diblock
copolymer.  In the framework of an Ornstein-Zernike approach for this
model system, repulsive \textit{inter}molecular interactions between
$A$ and $B$ species, at the \textit{monomer} level, are defined by the
direct pair correlation function containing hard-core intermolecular
repulsive interactions between like species, $C_{AA} = C_{BB} = C_0$,
and a sum of repulsive hard-core and tail potentials for
intermolecular interactions between monomers of unlike species,
$C_{AB} = C_{BA} = C_0 - \chi_{\mathit{eff}}/\rho$.  The effective
interaction parameter, $\chi_{\mathit{eff}}$, controls the amplitude
of microdomain scale concentration fluctuations and increases as the
system approaches its MST. Realistic diblock copolymer systems can be
mapped onto this simplified model, which has been extensively
investigated in the past \cite{blockschweizer}.  In the theoretical
coarse-graining approach presented here, the different chemical nature
of the two blocks is accounted for by the difference in their radii of
gyration, which is a function of the polymer local flexibility. This
assumption is well justified in our approach since the monomeric
structure is averaged out by the coarse-graining procedure, while
local flexibility enters through the block radii of gyration.

As a starting point in our derivation, we focus on the monomeric
quantities which are input to our coarse-grained description for a
diblock copolymer system, Eqs.\ (\ref{EQ:HBMS1}) and (\ref{EQ:HBMS}).
As a first approximation, we assume that all monomers comprising a
block are equivalent, so that each component in Eqs.\
(\ref{EQ:HBMS1}) and (\ref{EQ:HBMS}) represents a site-averaged
quantity. This is the conventional approximation adopted for treating
analytically integral equations for polymeric liquids, and becomes
correct when each block in the copolymer includes a number of monomers
large enough to minimize chain end effects. The approximation is
formally consistent with our coarse-graining theory, where physical
quantities at the monomer level are averaged over the monomer
distribution.

The monomer total pair correlation function
$h^{mm}_{\alpha\beta}(r)$, with $\alpha,\beta\in\{A,B\}$, in
reciprocal space is defined as the difference between the total static
structure factor and its intramolecular contribution as
\begin{equation}
\rho_{\alpha}\rho_{\beta}
h^{mm}_{\alpha\beta}(k) =
S^{mm}_{\alpha\beta}(k) - \rho\omega^{mm}_{\alpha\beta}(k) \, ,
\end{equation}
\noindent where the total pair correlation function obeys the
relationship
\begin{equation}
\rho^2h^{mm}_{\mathit{tot}}(k) =
\rho^2f^2h^{mm}_{AA}(k) + 2\rho^2f(1-f)h^{mm}_{AB}(k)
+ \rho^2(1-f)^2h^{mm}_{BB}(k)\, .
\end{equation}

We adopt here the thread model for the monomer-level description of
the polymer chain.  This model allows us to obtain analytical
equations for our coarse-grained description of the diblock copolymer
system.  In the thread model, the chain is treated as an infinite
thread of vanishing thickness, with hard-core monomer diameter
approaching zero, $d \to 0$, and a diverging segment number density in
the chain, $\rho_{\mathit{intra}} \to \infty$, while
$\rho_{\mathit{intra}} d^3$ remains finite.  The thread model yields a
good description of properties at the length scale of $R_g$,
including the presence of a correlation hole in the monomer pair
correlation function. It cannot account for the local fine structure
observed in the radial pair distribution function, $g(r) = 1 + h(r)$,
which is related to the presence of solvation shells due to monomer
hard-core interactions.  However, since our renormalized structures
are characterized by a size comparable to the block domain, it gives a
good representation of the coarse-grained structure for long chains
where the block size is larger than the monomer diameter.

\noindent  Each of the site-averaged components of the total static
structure factor contains contributions from density $S^{\rho}(k)$ and
concentration $S^{\phi}(k)$ fluctuations, and
can be expressed as a function of the density screening length as\cite{MDDYN}
\begin{align}
S^{mm}_{AA}(k) & =
\rho \frac{1 + k^2\xi_{\rho}^2/(1-f)}
{\left(1 + k^2\xi_{\rho}^2\right)/S^{\phi}(k)} \nonumber \\
S^{mm}_{AB}(k) & =
\rho\frac{1}
{\left(1 + k^2\xi_{\rho}^2\right)/S^{\phi}(k)} \nonumber \\
S^{mm}_{BB}(k) & =
\rho\frac{1 + k^2\xi_{\rho}^2/f}
{\left(1 + k^2\xi_{\rho}^2\right)/S^{\phi}(k)} \, .
\end{align}

The incompressible concentration structure factor is defined in
Leibler's mean-field approach as
\begin{equation}
\frac{N}{S^{\phi}(k)} = F(k) - 2N\chi_{\mathit{eff}} \, ,
\label{EQ:SINC}
\end{equation}

\noindent and diverges at the spinodal temperature, where
$\chi_{\mathit{eff}} = \chi_{\mathit{s}}=F(k^*)/(2N)$ with the
spinodal temperature defined as $N\chi_s \propto T_s^{-1}$. The
function  $F(k) = N\omega^{mm}_{\mathit{tot}}(k)/\omega(k)$ depends on
the intramolecular static structure factors
$\omega^{mm}_{\alpha\beta}(k)$ with $\alpha,\beta \in \{A,B\}$ through
the definition of $\omega(k)$  given by Eq.\ (\ref{EQ:omega}), and
\begin{equation}
\omega^{mm}_{\mathit{tot}}(k) =
\omega^{mm}_{AA}(k) + 2\omega^{mm}_{AB}(k) + \omega^{mm}_{BB}(k) \, .
\end{equation}

\noindent To take into account the phase stabilization due to
fluctuation effects, it is convenient to rewrite the incompressible
structure factor in the following approximate form \cite{FREDRICKSON},
\begin{equation}
N/S^{\phi}(k) = F(k) - 2N\chi_{\mathit{eff}} =
F(k) - 2N\chi_s + N/S^{\phi}(k^*) \, .
\label{EQ:SCOR}
\end{equation}

\noindent This expression takes into account the fact that when the
spinodal condition is fulfilled, the inverse concentration
contribution of the structure factor does not vanish: the disordered
phase is still present and eventually the system undergoes a
first-order phase transition.

With the purpose of obtaining an analytical expression for the
coarse-grained system, we introduce the Otha-Kawasaki approximation
\cite{OHTAK} given by $F(k) \approx (A/x + Bx - 2[AB]^{1/2}) +
2N\chi_s$, with $x = k^2R_g^2$. Here, $A = 3/[2f^2(1-f)^2]$ and $B =
1/[2f(1-f)]$.  At the peak position, defined as $x = x^* = [A/B]^{1/2}
= \{3/[f(1-f)]\}^{1/2}$, the contribution $(A/x + Bx - 2[AB]^{1/2})
\approx 0$, thus recovering the spinodal condition of $\chi_s =
F(k^*)/(2N)$. By inserting the Ohta-Kawasaki approximation into Eq.\
(\ref{EQ:SCOR}), the incompressible concentration fluctuation
collective structure factor reduces to
\begin{equation}
N/S^{\phi}(k) \approx A/x + Bx - 2[AB]^{1/2} + N/S^{\phi}(k^*) \, ,
\end{equation}

\noindent which after expanding into partial fractions, leads to the
following tractable analytical expression for the concentration
fluctuation contribution to the static structure factor \cite{MDDYN},
\begin{align}
S^{\phi}(k) & \approx 
\frac{6}{\sigma^2B}
\frac{k^2}{\left(k^2 + \xi_1^{-2} - \xi_2^{-2}\right)^2
+ 4\xi_1^{-2}\xi_2^{-2}} \, ,
\label{EQ:SKPHI}
\end{align}

\noindent characterized by two length scales $\xi_1 = \sigma
[2BS^\phi(k^*)/3]^{1/2}$ and $\xi_2 = \sigma \{2BN/[12(AB)^{1/2} -
3N/S^{\phi}(k^*)]\}^{1/2}$.  At the peak position, $k=k^* =
(A/B)^{1/4}/R_g$, the concentration structure factor behaves as
$S^{\phi}(k) = S^{\phi}(k^*)$.  In the small wave vector limit $(k \to
0)$, it increases as $S^{\phi}(k) \propto k^2$, in agreement with the
observed scaling behavior for homopolymer liquids.  Consistently, for
the large wave vector limit $(k \gg k^*)$, it tends to zero as
$k^{-2}$ since $S^{\phi}(k) = N/\left(Bx\right)$, thereby recovering
Leibler's scaling. The scaling with $k$ at large and small wave
vectors is also observed when homopolymer systems are investigated.
This is a characteristic feature of block copolymer systems,
signifying that at very large scales, as well as on very local scales,
fluctuations are independent of the effective repulsion between
monomers of unlike chemical nature.

The solution of Eqs.\ (\ref{EQ:HBMS1}) and (\ref{EQ:HBMS}) relies on
the definition of monomer-monomer and block-monomer intramolecular
form factors.  
The monomer-monomer form factors are well described by the
approximated Debye function,
\begin{align}
\omega^{mm}_{AA}(k) & \approx
\frac{2fN_A}{k^4R_{gA}^4}
\left[k^2R_{gA}^2 - 1 + e^{-k^2R_{gA}^2}\right]
\nonumber \\
\omega^{mm}_{BB}(k) & \approx
\frac{2(1-f)N_B}{k^4R_{gB}^4}
\left[k^2R_{gB}^2 - 1 + e^{-k^2R_{gB}^2}\right]
\nonumber \\
\omega^{mm}_{AB}(k) & \approx
\frac{N}{k^4R_{g}^4}
\left[k^2R_{gA}^2-1 + e^{-k^2R_{gA}^2}\right]
\left[k^2R_{gB}^2-1 + e^{-k^2R_{gB}^2}\right] \, ,
\label{EQ:DBYE}
\end{align}
which can be
conveniently simplified into their corresponding Pad\'e approximants.\cite{doiedw}.
In analogy with the center-of-mass monomer formalism
\cite{YAMAKAWA}, we approximate the block-monomer intramolecular structure factor in
reciprocal space by the following  Gaussian
distribution function
\begin{equation}
\omega^{bm}_{\alpha\beta}(k) = (1/2)N_{\beta}
e^{-k^2R_{g\alpha\beta}^2/6} \, ,
\label{EQ:WBMK}
\end{equation}

\noindent which includes the mean-squared radius of gyration,
describing the squared average distance of a monomer of type $\beta$
from the center of mass of an $\alpha$-type block,
\begin{equation}
R_{g\alpha\beta}^2 = \frac{1}{N_{\beta}}\sum^{N_{\beta}}_{i = 1}
\left(\vec{r}_{\beta_i} - \vec{R}_{b\alpha}\right)^2 \, .
\label{EQ:LDEF}
\end{equation}

\noindent In real space, $\omega^{bm}_{\alpha\beta}(r)$ is the generic
intramolecular distribution function for any one of the $N_{\beta}$
segments in a $\beta$-type block with respect to the center of mass of
an $\alpha$-type block.  Finally, we define the intramolecular total
block-monomer structure factor as
\begin{equation}
\omega^{bm}_{\mathit{tot}}(k) =
\omega^{bm}_{AA}(k) + \omega^{bm}_{AB}(k)
+ \omega^{bm}_{BA}(k) + \omega^{bm}_{BB}(k) \, ,
\label{EQ:WBMT}
\end{equation}

\noindent which in the $k\to0$ regime yields
$\omega^{bm}_{\mathit{tot}}(0)= 1/2\left[f + (1-f)\right]N +
1/2\left[f + (1-f)\right]N=N$.

While Eq.\ (\ref{EQ:WBMK}) is a well-known expression when it applies
to the distribution of monomers around the center-of-mass of an
unperturbed homopolymer chain \cite{YAMAKAWA}, its extension to
diblock copolymers is novel. When tested against united atom
simulation data (see Fig. \ref{FG:WBMK} and the discussion of Section
\ref{SX:COMP}), these analytical expressions are fairly accurate for
both symmetric and asymmetric diblock copolymers, while their simple
Gaussian form allows us to derive analytical equations for the block
total correlation functions in real space.

\section{Block coarse-grained  description in reciprocal space, and isothermal compressibility}
\label{SX:SYMM}
\noindent In the large-$k$ regime, which is of interest in block
copolymer melts due to the finite size of the microphase separation,
the ratios of intramolecular structure factors follow the relations
$\omega^{mm}_{AA}(k)/\omega(k) \approx k^2\sigma^2/[12(1-f)]$,
$\omega^{mm}_{BB}(k)/\omega(k) \approx k^2\sigma^2/[12f]$, and
$\omega^{mm}_{AB}(k)/\omega(k) \approx 1/[4N f(1-f)]$.  By enforcing
the approximation that $\omega^{mm}_{AB}(k) \ll \{\omega^{mm}_{AA}(k),
\omega^{mm}_{BB}(k)\}$, which is justified by the fact that
$\omega^{mm}_{AB}(k)/\omega(k)\propto N^{-1} $ and vanishes for long
polymer chains, $N \rightarrow \infty$, the total block-monomer and
block-block correlation functions simplify. Including these
approximations into the monomer-monomer and block-monomer structure
factors reduce Eqs.\ (\ref{EQ:HBMS1}) to the analytical general forms
\begin{gather}
\begin{split}
h^{bm}_{\alpha A}(k)/2 &=
\left[\frac{\omega^{bm}_{\alpha A}(k)+\omega^{bm}_{\alpha B}(k)}
{\omega^{mm}_{\mathit{tot}}(k)}\right]
h^{\rho}(k)
+
\left[\frac{f^{-1}\omega^{bm}_{\alpha A}(k)-(1-f)^{-1}
\omega^{bm}_{\alpha B}(k)}
{\omega^{mm}_{\mathit{tot}}(k)}\right]
(1-f)\Delta h^{\phi}(k) \, , \\
h^{bm}_{\alpha B}(k)/2 &=
\left[\frac{\omega^{bm}_{\alpha A}(k)+\omega^{bm}_{\alpha B}(k)}
{\omega^{mm}_{\mathit{tot}}(k)}\right]
h^{\rho}(k)
-
\left[\frac{f^{-1}\omega^{bm}_{\alpha A}(k)-(1-f)^{-1}
\omega^{bm}_{\alpha B}(k)}
{\omega^{mm}_{\mathit{tot}}(k)}\right]
f\Delta h^{\phi}(k) \, ,
\label{EQ:hBMkup}
\end{split}
\end{gather}

\noindent with $\alpha \in \{A,B\}$.

Following the same procedure, the block-block total correlation
functions in reciprocal space, Eqs.\ (\ref{EQ:HBMS}), reduce to
\begin{gather}
\begin{split}
h^{bb}_{\alpha \beta}(k)/4  = &
\frac{\left[\omega^{bm}_{\alpha A}(k)
+ \omega^{bm}_{\alpha B}(k)\right]
\left[\omega^{bm}_{\beta A}(k)
+ \omega^{bm}_{\beta B}(k)\right]}
{\left[\omega^{mm}_{\mathit{tot}}(k)\right]^2}
h^{\rho}(k) \\
+ & \frac{\left[f^{-1}\omega^{bm}_{\alpha A}(k)
- (1-f)^{-1}\omega^{bm}_{\alpha B}(k)\right]
\left[f^{-1}\omega^{bm}_{\beta A}(k)
- (1-f)^{-1}\omega^{bm}_{\beta B}(k)\right]}
{\left[\omega^{mm}_{\mathit{tot}}(k)\right]^2}
 f(1-f)\Delta h^{\phi}(k)\, ,
\label{EQ:HBBAB}
\end{split}
\end{gather}

\noindent where the density contribution $h^{\rho}(k)$ is identical to
the monomer total correlation function for a homopolymer chain
\cite{PRISM}
$ h^{\rho}=4 \pi \xi_{\rho}'
\left[ {\xi_{\rho}^2}/(1+\xi_{\rho}^2 k^2) -
\xi_{c}^2 /(1+\xi_{c}^2 k^2)\right]$,
and the concentration fluctuation contribution  at some
thermal state point $N\chi_{\mathit{eff}}\propto T^{-1}$, having as a
reference the athermal state $N\chi_{\mathit{eff}} = 0$, is derived
from Eq.\ (\ref{EQ:SKPHI}) as
$\Delta h^{\phi}(k) =
h^{\phi}_{N\chi_{\mathit{eff}}}(k) - h^{\phi}_0(k)$
with
$h^{\phi}(k) =
[f(1-f)]^{-1}\rho^{-1} S^{\phi}(k)$.

The total block-monomer intermolecular pair correlation
function reads
\begin{gather}
\begin{split}
h^{bm}_{\mathit{tot}}(k) & =
  fh^{bm}_{AA}(k)/2
+ (1-f)h^{bm}_{AB}(k)/2
+ fh^{bm}_{BA}(k)/2
+ (1-f)h^{bm}_{BB}(k)/2 \\
& =
\left[\frac{\omega^{bm}_{\mathit{tot}}(k)}
{\omega^{mm}_{\mathit{tot}}(k)}\right]h^{mm}(k) \, ,
\label{EQ:HBMTO}
\end{split}
\end{gather}

\noindent where the contribution from concentration fluctuations
rigorously vanishes, as is the case for the monomer level description
of a diblock copolymer melt.  The total block-block intermolecular
pair correlation function is given by
\begin{eqnarray}
h^{bb}_{\mathit{tot}}(k) & = &
h^{bb}_{AA}/4 + h^{bb}_{AB}/2 +
h^{bb}_{BB}/4
= \left[\frac{\omega^{bm}_{\mathit{tot}}(k)}
{\omega^{mm}_{\mathit{tot}}(k)}\right]^2
h^{\rho}(k) \nonumber
\\
& + & \left[
\frac{f^{-1}\left(\omega^{bm}_{AA}(k)
+ \omega^{bm}_{BA}(k)\right)
- (1-f)^{-1}\left(\omega^{bm}_{AB}(k)
+ \omega^{bm}_{BB}(k)\right)}
{\omega^{mm}_{\mathit{tot}}(k)}\right]^2 \Delta h^{\phi}(k) \, .
\label{EQ:HBTO}
\end{eqnarray}

\noindent When compositionally asymmetric diblocks are investigated,
the concentration fluctuation contribution to Eq.\ (\ref{EQ:HBTO})
does not vanish, but instead provides a small correction to the
density fluctuation part. However, under athermal conditions or in the
$k \rightarrow 0$ limit, only density fluctuations are relevant since
the concentration fluctuation contribution vanishes in a manner
consistent with the monomer level description.  It is worth noting
that in the limit of a block approaching the size of the polymeric
molecule, and in the limit of blocks of identical length (see Section
\ref{AP:SYMM} of the Appendix), Eq.\ (\ref{EQ:HBTO}) recovers the
homopolymer expression for the molecular center-of-mass total pair
correlation function, with concentration fluctuations strictly
vanishing.

As a test of our formalism, we present in Section \ref{SX:COMP}
a comparison of Eqs. \ (\ref{EQ:hBMkup}), (\ref{EQ:HBBAB}) and Eqs.\
(\ref{EQ:HBMTO}), (\ref{EQ:HBTO}) against simulation data in the
athermal regime. All equations show good agreement with simulations
for both compositionally symmetric and asymmetric diblock copolymers
(see also Figs. \ref{FG:HBMaK} and \ref{FG:HBBK}), thus supporting 
the validity of our procedure.

Finally, starting from Eqs.\ (\ref{EQ:HBMTO}) and (\ref{EQ:HBTO}),
we calculate the isothermal
compressibility of the system. Since the latter is a bulk property, it
has to be independent of the level of coarse-graining adopted in the
description of molecules in the liquid.  The isothermal
compressibility of the system coarse-grained at the block-monomer
level, $\kappa_T=S_{\mathit{tot}}^{bm}(k\to0)/(\rho k_BT)$, is
obtained from the matricial definition $\mathbf{S}^{bm}(k) =
\mathbf{\Omega}^{bm}(k) + \mathbf{H}^{bm}(k)$, after taking the
$k\to0$ limit and adimensionalizing the static structure factor.  Each
contribution is given by the relation $S^{bm}_{\alpha \beta}(k\to0) =
N_{\beta}(\xi_{\rho}^2/\xi_c^2)/2$, which yields
$S^{bm}_{\mathit{tot}}(k\to0) = S^{mm}_{\mathit{tot}}(k\to0) =
N(\xi_{\rho}^2/\xi_c^2)$.  In an analogous way, the compressibility of
the system coarse-grained at the block-block level is calculated from
the relation  $\kappa_T=S^{bb}(k\to0)/(\rho k_BT)$, and it is obtained
from the matricial definition $\mathbf{S}^{bb}(k) =
\mathbf{\Omega}^{bb}(k) + \mathbf{H}^{bb}(k)$, where $S^{bb}_{\alpha
\beta}(k\to0) = (\xi_{\rho}^2/\xi_c^2)/2$.  Since
$NS^{bb}_{\mathit{tot}}(k\to0)/2 =
S^{bm}_{\mathit{tot}}(k\to0)=S^{mm}_{\mathit{tot}}(k\to0) =
N(\xi_{\rho}^2/\xi_c^2)$, this result is consistent with our prior
findings obtained when coarse-graining homopolymer melts at the
center-of-mass level, validating the feature that bulk properties,
such as $\kappa_T$, are independent of the fundamental unit (or frame
of reference) chosen to represent the system.

Reproducing the isothermal compressibility of the system, after performing a coarse-graining procedure, is an important test of the latter. Due to the nature of the coarse-graining procedure, effective coarse-grained potentials derived from pair distribution functions are softer than their microscopic counterparts. In fact, while real units, such as chain monomers, cannot physically superimpose, auxiliary sites can and the potential at contact is finite. For this reason, the occurrence of small errors in the evaluation of the potential, which is often the case for numerically evaluated coarse-grained potentials, yields liquids that are too compressible. This shortcoming is eliminated in systems for which coarse-grained total correlation functions can be evaluated  analytically, as it is in our case.

\section{Analytical block-level description in real space}
\label{SX:SYMMR}
\subsection{Block-monomer total correlation function}
\noindent For  a structurally and interaction symmetric diblock
copolymer, the total pair correlation functions for block-monomer and
block-block terms in real space can be expressed analytically through
a simple Fourier transform. The block-monomer expression separates
into density and concentration fluctuation contributions as
\begin{gather}
\begin{split}
h^{bm}_{\alpha \beta}(r) & =
{h^{bm,\rho}_{\alpha \beta}}(r)
+ {\Delta h^{bm,\phi}_{\alpha \beta}}(r) \, ,
\label{EQ:Hbmdir}
\end{split}
\end{gather}

\noindent with $\alpha, \beta \in \{A, B\}$. The density fluctuation
contribution is represented explicitly by the relations
\begin{gather}
\begin{split}
{h^{bm,\rho}_{\alpha \alpha}}(r) & =
fI^{\rho}(r,R_{g_{\alpha A}}) + (1-f)I^{\rho}(r,R_{g_{\alpha B}})
= {h^{bm,\rho}_{\alpha \beta}}(r) \, ,
\end{split}
\end{gather}

\noindent where for compactness, we introduce the auxiliary
function, $I^{\rho}(r,R)$, defined by Eq.\ (\ref{EQ:Hrho})
of the Appendix.  More specifically, Eq.\
(\ref{EQ:Hrho}) represents the density fluctuation contribution for
one block, and is identical in form to the expression derived in our
previous work for the center-of-mass-monomer total correlation
function in homopolymer melts, coarse-grained at the polymer
center-of-mass level \cite{MLTEX}. 

The concentration fluctuation contribution in real space is given by
the relations
\begin{gather}
\begin{split}
{f\Delta h^{bm,\phi}_{AA}}(r) & =
   \Delta I^{\phi}(r,R_{gAA})
 - \Delta I^{\phi}(r,R_{gAB}) =
 -{(1-f)\Delta h^{bm,\phi}_{AB}}(r)\, ,\\
{(1-f)\Delta h^{bm,\phi}_{BB}}(r) & =
  \Delta I^{\phi}(r,R_{gBB})
 - \Delta I^{\phi}(r,R_{gBA}) = -{f \Delta h^{bm,\phi}_{BA}}(r)\, ,
 \label{EQ:Hphitutti}
\end{split}
\end{gather}

\noindent where each term is defined as the difference in the response
of the concentration fluctuation contribution between some thermal
state ($N\chi_{\mathit{eff}}\propto T^{-1}$) and the reference
athermal state ($N\chi_{\mathit{eff}}=0$), as  $\Delta I^{\phi}
\left(r, R \right)= I^{\phi}_{N\chi_{\mathit{eff}}} \left(r, R \right)
- I^{\phi}_0 \left(r, R \right)$, with $I^{\phi} \left(r, R \right)$
defined by Eq.\ (\ref{EQ:gener}) of the Appendix. 

In the microscopic, small $r/R$ regime, the concentration fluctuation
contribution in Eq.\ (\ref{EQ:Hphitutti}) reduces to
\begin{gather}
\begin{split}
 \Delta I^{\phi}(r,R) & \approx
\sqrt{\frac{6}{\pi}} f^2(1-f)^2
\frac{\xi_{\rho}'}{2 R}
e^{-3r^2/(2R^2)} N\chi_{\mathit{eff}} \, .
\label{pena}
\end{split}
\end{gather}

\noindent This is the regime most relevant for block copolymer liquids
approaching their phase transition, since the microphase separation
transition is characterized by segregation on spatial scales close in
magnitude to the polymer radius of gyration.  
The temperature dependence enters Eq.(\ref{EQ:Hphitutti}) through the 
$\chi_{\mathit{eff}}$ parameter in Eq.(\ref{pena}), evaluated at the reference athermal and thermal states.
In this way, at high temperatures, i.e.\ far from the phase transition, density
fluctuations are dominant over concentration fluctuations, and the
total correlation function for diblock copolymer liquids recovers that
of the homopolymer.  

In proximity of the phase transition and for long polymeric chains, Eqs.\
(\ref{EQ:Hphitutti}) further simplify with the cross terms becoming negligible,  while self terms
yield the main contribution,
\begin{gather}
\begin{split}
{f\Delta h^{bm,\phi}_{AA}}(r) & =
-{(1-f)\Delta h^{bm,\phi}_{AB}}(r) \approx
\sqrt{\frac{6}{\pi}} f^2(1-f)^2
\frac{\xi_{\rho}'}{2 R_{gAA}}
e^{-3r^2/(2R_{gAA}^2)} N\chi_{\mathit{eff}} , \\
 {(1-f)\Delta h^{bm,\phi}_{BB}}(r) & =
  -{f \Delta h^{bm,\phi}_{BA}}(r) \approx
\sqrt{\frac{6}{\pi}} f^2(1-f)^2
\frac{\xi_{\rho}'}{2 R_{gBB}}
e^{-3r^2/(2R_{gBB}^2)} N\chi_{\mathit{eff}} .
 \label{EQ:Hphituttiaaprox}
\end{split}
\end{gather}

\noindent Here, $R_{g\alpha \beta}$ is the average distance of a
monomer of type $\beta$ from the center-of-mass of the block of type
$\alpha$, as defined in Eq.\ (\ref{EQ:LDEF}).  A numerical study of
these approximated expressions shows that neglecting terms with the
most ``cross'' character is a reasonable approximation in real space,
valid under different temperature limits and even when the system is
asymmetric.

A measure of the physical clustering with temperature
 is given by the parameter
\begin{eqnarray}
\Delta g^{bm}(r)=
g_{AA}^{bm}(r)+g_{BB}^{bm}(r)-g_{AB}^{bm}(r)-g_{BA}^{bm}(r)\, ,
\label{deltag}
\end{eqnarray}

\noindent where $g_{\alpha \beta}^{bm}(r)=1+h_{\alpha \beta}^{bm}(r)$.
Since the number of monomers of type $\beta$ included within a sphere
of radius $r'$ from the center-of-mass of block $\alpha$ is given by
\begin{eqnarray}
n^{bm}_{\hphantom{\alpha}\beta}(r') =4 \pi\rho_{\beta}
\int_0^{r'} r^2 g_{\alpha \beta}^{bm} (r) dr \ ,
\end{eqnarray}

\noindent with $\alpha, \beta \in \{A,B\}$, Eq.\ (\ref{deltag})
represents a measure of the clustering of monomers around blocks of
like species. The density fluctuation contribution to Eq.\
(\ref{deltag}) exactly cancels, while the concentration fluctuation
contribution increases with decreasing temperature (increasing $N\chi_{\mathit{eff}}$). At contact
($r\to0$), we obtain
\begin{equation}
\Delta g^{bm}(0)\approx
\sqrt{\frac{3}{2\pi}} f(1-f) \xi'_\rho
\left(R_{gAA}^{-1}+R_{gBB}^{-1} \right) N\chi_{\mathit{eff}}\, .
\end{equation}

\noindent In the athermal limit block-monomer clustering due to concentration
fluctuations vanishes, and $\Delta g^{bm}(0)=0$. At lower temperatures
clustering of like species increases, with the leading factor being proportional to the ratio
$\xi_{\rho}'/R_{g\alpha \beta}$, which control the strength of finite-size coupling
of microdomains ($R_{g\alpha \beta}$) and local ($\xi_{\rho}'$) correlations.

\subsection{Block-block total correlation function}
\noindent The block-block intermolecular total pair correlation
functions can be solved in an analogous way to give in real space,
\begin{align}
h^{bb}_{\alpha \beta}(r) &= {h^{bb,\rho}_{\alpha \beta}}(r)
+ \Delta{h^{bb,\phi}_{\alpha \beta}}(r)\, ,
\end{align}

\noindent where the separation between density and concentration
contributions is conserved.  The density fluctuation contribution is
given by
\begin{eqnarray}
{h^{bb,\rho}_{\alpha \beta}}(r)& = &
f^2 J^{\rho}  \left(r, R_{\alpha A \beta A}\right)
+ f(1-f) J^{\rho} \left(r, R_{\alpha A \beta B}\right) \nonumber \\
& + & f(1-f) J^{\rho} \left(r, R_{\alpha B \beta A}\right)
+ (1-f)^2 J^{\rho}\left(r, R_{\alpha B \beta B}\right) \, ,
\label{EQ:Hbbrhotutti}
\end{eqnarray}

\noindent with $\alpha, \beta\in\{A,B\}$,  and $J^{\rho}(r, R)$ defined
by Eq.\ (\ref{EQ:GEN}) of the Appendix. The distance 
$\left(R_{\alpha\beta\gamma\delta}\right)^2 =
[\left(R_{g\alpha\beta}\right)^2 +
\left(R_{g\gamma\delta}\right)^2]/2$, where 
$R_{g\alpha \beta}$ is the average distance of a
monomer of type $\beta$ from the center-of-mass of the block of type
$\alpha$, as defined in Eq.\ (\ref{EQ:LDEF}). We note that Eq.\
(\ref{EQ:GEN}) was previously derived by us, in the context of
coarse-graining a homopolymer liquid at the center-of-mass level
\cite{YAPRL,MLTEX,JOPCM}.  This term represents the total pair
correlation function for a liquid of interacting soft colloidal
particles, centered on the spatial position of the polymer center of
mass. This simple analytical expression reproduces well data from
united atom molecular dynamics simulations of polymer melts.

The concentration fluctuation contribution is given by the general
equation
\begin{eqnarray}
\Delta {h^{bb,\phi}_{\alpha \beta}}(r) & = &
  \Delta J^{\phi}\left(r, R_{\alpha A \beta A}\right)
-\Delta J^{\phi}\left(r, R_{\alpha A \beta B}\right) \nonumber \\
&-&\Delta J^{\phi}\left(r, R_{\alpha B \beta A}\right)
+ \Delta J^{\phi}\left(r, R_{\alpha B \beta B}\right) \, ,
\label{EQ:Hbbphitutti}
\end{eqnarray}

\noindent where we define $\Delta J^{\phi} \left(r, R \right) =
J^{\phi}_{N\chi_{\mathit{eff}}} \left(r, R \right) - J^{\phi}_0
\left(r, R \right)$ and the auxiliary function $J^{\phi}(r, R)$ by
Eq.\ (\ref{EQ:jei}) of the Appendix.  In the small $r/R_g$ regime of
interest here, the concentration fluctuation contribution simplifies,
yielding for the generic contribution in Eq.\ (\ref{EQ:Hbbphitutti})
the following simplified expression
\begin{eqnarray}
\Delta J^{\phi}(r, R) \approx
f^2 (1-f)^2   \sqrt{\frac{3}{\pi}}
\frac{\xi_{\rho}'}{R} e^{-3r^2/(4R^2)}
\left(2+ 3 \frac{\xi_c^2}{R^2}\right) N\chi_{\mathit{eff}} \ .
\end{eqnarray}

For long polymeric chains in general the cross statistical distances are larger than the self ones,
$R_{g \alpha \beta } >>  R_{g \alpha \alpha}$, 
and Eqs.\ (\ref{EQ:Hbbphitutti}) simplify to

\begin{eqnarray}
\Delta {h^{bb,\phi}_{AA}}(r) & \approx
& f^2 (1-f)^2   \sqrt{\frac{3}{\pi}}
\frac{\xi_{\rho}'}{R_{gAA}} e^{-3r^2/(4R_{gAA}^2)}
\left(2+ 3 \frac{\xi_c^2}{R_{gAA}^2}\right) N\chi_{\mathit{eff}}
\nonumber\\
\Delta {h^{bb,\phi}_{AB}}(r) & = &
 \Delta {h^{bb,\phi}_{BA}}(r) \approx - f^2 (1-f)^2
\sqrt{\frac{3}{\pi}}
\frac{\xi_{\rho}'}{\sqrt{(R_{gAA}^2+R_{gBB}^2)/2}}
e^{-3r^2/[2(R_{gAA}^2+R_{gBB}^2)]} \nonumber\\
& &\left(2+ 3 \frac{\xi_c^2}{(R_{gAA}^2+R_{gBB}^2)/2}\right)
N\chi_{\mathit{eff}} \nonumber\\
\Delta {h^{bb,\phi}_{BB}}(r) & \approx & f^2 (1-f)^2
\sqrt{\frac{3}{\pi}}
\frac{\xi_{\rho}'}{R_{gBB}} e^{-3r^2/(4R_{gBB}^2)}
\left(2+ 3 \frac{\xi_c^2}{R_{gBB}^2}\right) N\chi_{\mathit{eff}} \, .
\end{eqnarray}

\noindent As with the block-monomer functions, numerical tests show that neglecting those ``cross''
terms is a reasonable approximation that holds under different
temperature limits and even when the system is asymmetric.

In analogy to the block-monomer development, to study the buildup of
concentration fluctuations we define the parameter $\Delta g^{bb}(r)$,
which represents a measure of the physical clustering with
temperature of blocks of like species, as
\begin{eqnarray}
\Delta g^{bb}(r)=
g_{AA}^{bb}(r)+g_{BB}^{bb}(r)-2g_{AB}^{bb}(r)\ .
\label{deltagbb}
\end{eqnarray}

\noindent The number of $\beta$-type blocks included within a sphere
of radius $r'$ from the center-of-mass of block $\alpha$, is given by
\begin{eqnarray}
n^{bb}_{\alpha}(r')=4 \pi\rho_{b\beta}
\int_0^{r'} r^2 g_{\alpha \beta}^{bb} (r) dr
+ \delta_{\alpha \beta} \ ,
\end{eqnarray}

\noindent with $\alpha,\beta \in \{A,B\}$. Clustering due to
concentration fluctuations increases with decreasing temperature,
while density fluctuations provide a contribution constant with
temperature, which is a consequence of the asymmetry in diblock
composition and vanishes for compositionally symmetric diblocks. The
scaling with degree of polymerization of the function  $\Delta
g^{bb,\phi}(r) \propto \sqrt{N} / S(k^*)$ depends on how far the
system is from its microphase separation transition. At temperatures
higher than the order-disorder temperature ($T \gg T_{\mathit{ODT}}$),
we find that $\Delta g^{bb,\phi}(r) \propto N^{-1/2}$. At the
transition temperature ($T \approx T_{\mathit{ODT}}$), $\Delta
g^{bb,\phi}(r) \propto N^{-5/6}$, while in the low temperature regime
($T \ll T_{\mathit{ODT}}$),  $\Delta g^{bb,\phi}(r) \propto N^{-3/2}$.
With respect to the monomer-block coarse-graining, $\Delta g^{bb}(r)$ contains
a second term where the leading factor is proportional to the lengthscale ratio 
$\xi_c/R_g$.

\section{Coarse-graining of diblock copolymers at the center-of-mass level}
\label{SX:DCOM}
\subsection{Reciprocal space representation}
\noindent In this section, we describe a diblock copolymer melt
coarse-grained at the center-of-mass level.  Information at this
resolution is heavily averaged since the length scale of
coarse-graining, $R_g$, is larger than the block size. However, it is
still useful to consider this description since it characterizes
phenomena on the length scale of the polymer radius of gyration, and
establishes a formal bridge of the theory presented here with previous
approaches to coarse grain homopolymer melts and their mixtures at the
center-of-mass level \cite{YAPRL,MLTEX,JOPCM,BLNDS}.

To derive a coarse-graining procedure that maps block copolymer chains
onto soft colloidal particles, centered on the coordinates of the
polymer center of mass, the Ornstein-Zernike matricial relation is
first specialized to include ``auxiliary'' center-of-mass sites.  
Here intramolecular structure factor matrix contains the matrix of real sites correlation defined before,
as well as $\mathbf{\Omega}^{cm}_{\alpha}=\rho_{ch} \omega^{cm}_{\alpha}$ with the number
density of chain $\rho_{ch}=\rho/N$,
and $\mathbf{\Omega}^{cc}=\rho_{ch}$. The matrix of the total pair correlation functions contains 
the elements $\mathbf{H}^{cm}_{\alpha}=\rho_{ch}^2 N_{\alpha} h_{\alpha}^{cm}$ and
$\mathbf{H}^{cc}=\rho_{ch}^2 h^{cc}$.
The intermolecular direct correlation function matrix follows the
usual assumption that there is neither a correlation between auxiliary
sites nor with any other type of site, such that the only
non-vanishing element is the contribution from $\mathbf{C}^{mm}(k)$.
Following analogous approximations and the analytical development of
Section \ref{AP:SYMM}, we arrive to a representation
of the relations cited above that rigorously decouples density and
concentration fluctuations.  This arrangement is given by the
following set of expressions for the center-of-mass-monomer total
correlation functions,
\begin{gather}
\begin{split}
h^{cm}_{\hphantom{A}A}(k) &=
\left[\frac{\omega^{cm}_{\mathit{tot}}(k)}
{\omega^{mm}_{\mathit{tot}}(k)}\right]
h^{\rho}(k)+
\left[\frac{f^{-1}\omega^{cm}_{\hphantom{A}A}(k)-(1-f)^{-1}
\omega^{cm}_{\hphantom{A}B}(k)}
{\omega^{mm}_{\mathit{tot}}(k)}\right]
(1-f)\Delta h^{\phi}(k) \, , \\
h^{cm}_{\hphantom{B}B}(k) &=
\left[\frac{\omega^{cm}_{\mathit{tot}}(k)}
{\omega^{mm}_{\mathit{tot}}(k)}\right]
h^{\rho}(k) -
\left[\frac{f^{-1}\omega^{cm}_{\hphantom{B}A}(k)
-(1-f)^{-1}\omega^{cm}_{\hphantom{B}B}(k)
}
{\omega^{mm}_{\mathit{tot}}(k)}\right]
f\Delta h^{\phi}(k) \, ,
\label{EQ:HCMSL}
\end{split}
\end{gather}

\noindent and for the center-of-mass total correlation function,
\begin{multline}
h^{cc}(k) =
\left[\frac{\omega^{cm}_{\mathit{tot}}(k)}
{\omega^{mm}_{\mathit{tot}}(k)}\right]^2
h^{\rho}(k)
+ \left[\frac{f^{-1}\omega^{cm}_{\hphantom{c}A}(k)
-(1-f)^{-1}\omega^{cm}_{\hphantom{c}B}(k)}
{\omega^{mm}_{\mathit{tot}}(k)}\right]^2
f(1-f)\Delta h^{\phi}(k)\, ,
\label{EQ:HCSL}
\end{multline}

\noindent where we include the relation
\begin{equation}
\omega^{cm}_{\mathit{tot}}(k) =
\omega^{cm}_{\hphantom{c}A}(k) +
\omega^{cm}_{\hphantom{c}B}(k)\, .
\label{EQ:WCMT}
\end{equation}

%\noindent While formally exact, Eq.\ (\ref{EQ:WCMT}) becomes
%approximate when representing the individual contributions by Gaussian
%form factors.  
For the center-of-mass monomer intramolecular
correlation function we start from the definition  
\cite{YAMAKAWA} $\omega^{cm}_{\mathit{tot}}(k) =
Ne^{-k^2R_g^2/6}$, which leads to
\begin{equation}
\omega^{cm}_{\hphantom{c}\alpha}(k) = N_{\alpha}
e^{-k^2R_{gc\alpha}^2/6}\, ,
\label{EQ:WCMK}
\end{equation}

\noindent with
\begin{equation}
R_{gc\alpha}^2 = \frac{1}{N_\alpha} \sum^{N_\alpha}_{i=1}
\left(\vec{r}_{\alpha_i}-\vec{R}_c\right)^2\, ,
\end{equation}

\noindent representing the radius of gyration involving $N_{\alpha}$
segments and the molecular center-of-mass coordinate, $\vec{R}_c$.  A
justification for Eq.\ (\ref{EQ:WCMK}) can be found by performing the
small-$k$ expansion of Eq.\ (\ref{EQ:WCMT}).   In the athermal limit, Eqs.\
(\ref{EQ:HCMSL}) and (\ref{EQ:HCSL}) correctly recover the homopolymer
melt expressions \cite{YAPRL,MLTEX}.

In the case of a structurally and compositionally symmetric system,
where $\omega^{mm}_{AA}(k) = \omega^{mm}_{BB}(k)$ and
$\omega^{cm}_{\hphantom{c}A}(k) = \omega^{cm}_{\hphantom{c}B}(k)$, the
equations further simplify recovering the known relation for
homopolymer melts \cite{YAPRL,MLTEX},
$h^{cm}(k)=[\omega^{cm}(k) /\omega^{mm}(k)]h^{\rho}(k)$ and
$ h^{cc}(k)= [\omega^{cm}(k)/\omega^{mm}(k)]^2 h^{\rho}(k)$,
as expected.

The sum of domain-resolved contributions for the
intermolecular center-of-mass-monomer total correlation functions
yields an expression analogous to the one obtained for polymer melts, where
concentration fluctuations terms rigorously vanish, namely,
\begin{gather}
\begin{split}
h^{cm}_{\mathit{tot}}(k) =
fh^{cm}_{\hphantom{A}A}(k)
+ (1-f)h^{cm}_{\hphantom{A}B}(k) =
\left[\frac{\omega^{cm}_{\mathit{tot}}(k)}
{\omega^{mm}_{\mathit{tot}}(k)}\right]h^{mm}(k) \, .
\end{split}
\end{gather}

Finally, the isothermal compressibility of the system at the present
coarse-grained level, $\kappa_T=S^{cm}(k\to0)/(\rho k_BT)$, can be
obtained from the matricial definition $\mathbf{S}^{cm}(k) =
\mathbf{\Omega}^{cm}(k) + \mathbf{H}^{cm}(k)$, after taking the
$k\to0$ limit and adimensionalizing the static structure factor.  The
respective contributions are given by the relations
\begin{gather}
\begin{split}
S^{cm}_{\hphantom{A}A}(k\to0) &= fN(\xi_{\rho}^2/\xi_c^2) \, ,\\
S^{cm}_{\hphantom{A}B}(k\to0) &= (1-f)N(\xi_{\rho}^2/\xi_c^2) \, ,
\end{split}
\end{gather}

\noindent which yields after summing each contribution, that
$S^{cm}_{\mathit{tot}}(k\to0) = S^{mm}_{\mathit{tot}}(k\to0) =
N(\xi_{\rho}^2/\xi_c^2)$, a result consistent with the expression for
the liquid compressibility derived from coarse-graining homopolymer
fluids at the center-of-mass level \cite{YAPRL,MLTEX,JOPCM,BLNDS}.

\subsection{Real space representation}
\noindent An analogous representation is afforded in real space, where
density ($\rho$) and concentration ($\phi$) fluctuation contributions
separate as
\begin{gather}
\begin{split}
h^{cm}_{\hphantom{A}\alpha}(r) & =
{h^{cm,\rho}_{\hphantom{A}\alpha}}(r)
+ {\Delta h^{cm,\phi}_{\hphantom{A}\alpha}}(r)\, , \\
h^{cc}(r) & = {h^{cc,\rho}}(r)
+ \Delta{h^{cc,\phi}}(r)
\, ,
\end{split}
\end{gather}

\noindent with $\alpha \in \{A,B\}$.

Since the integrands needed for the real space representation are
identical in form to those presented in Section \ref{SX:SYMMR}, we
simply cite the solution in terms of the respective functions.  The
density fluctuation contribution is represented explicitly by the
relations
\begin{gather}
\begin{split}
{h^{cm,\rho}_{\hphantom{A}A}}(r) & =
fI^{\rho}(r,R_{g_{cA}}) + (1-f)I^{\rho}(r,R_{g_{cB}})
= {h^{cm,\rho}_{\hphantom{A}B}}(r) \, , \\
{h^{cc,\rho}}(r) & = J^{\rho}(r,R_g)\, ,
\label{EQ:HCCG}
\end{split}
\end{gather}

\noindent In the limit of $N\to\infty$, the exact solution for the
density fluctuation contribution,
${h^{cm,\rho}_{\hphantom{A}\alpha}}(r)$ and $h^{cc,\rho}(r)$, can be
conveniently approximated \cite{YAPRL,MLTEX} by
\begin{equation}
\label{EQ:HCMA}
h^{cm,\rho}(\tilde{r},\tilde{\xi}_{\rho}) \approx
-\frac{3}{2}\sqrt{\frac{6}{\pi}}
\tilde{\xi}_{\rho}\left(1+\sqrt{2}\tilde{\xi}_{\rho}\right)
\left[1+
\mathcal{O}\left(\tilde{\xi}_{\rho}^2,\tilde{r}^2\right)\right]
e^{-3\tilde{r}^2/2}\, ,
\end{equation}

\noindent and
\begin{equation}
h^{cc,\rho}\left(\tilde{r},\tilde{\xi}_{\rho}\right)
\approx
- \frac{39}{16}\sqrt{\frac{3}{\pi}}
\tilde{\xi}_{\rho}\left(1 + \sqrt{2}\tilde{\xi}_{\rho}\right)
\left[1 - \frac{9\tilde{r}^2}{26} +
\mathcal{O}\left(\tilde{\xi}_{\rho}^2,\tilde{r}^4\right)\right]
e^{-3\tilde{r}^2/4} \, ,
\end{equation}

\noindent where $\tilde{\xi}_{\rho}=\xi_{\rho}/R_g$ is the rescaled
density fluctuation correlation length scale and $\tilde{r}=r/R_g$ is
the rescaled distance between intermolecular center-of-mass sites.

The concentration fluctuation contribution denotes, as before, a
difference between two thermodynamic conditions for the system under
study, more specifically, between a thermal state
($N\chi_{\mathit{eff}}\propto T^{-1}$) and the athermal reference
state ($N\chi_{\mathit{eff}}=0$), yielding the expressions
\begin{gather}
\begin{split}
{f\Delta h^{cm,\phi}_{\hphantom{A}A}}(r) & =
   \Delta I^{\phi}(r,R_{g_{cA}})
 - \Delta I^{\phi}(r,R_{g_{cB}})
 = -{(1-f)h^{cm,\phi}_{\hphantom{A}B}}(r)\, , \\
 \Delta{h^{cc,\phi}}(r)  & =
\Delta J^{\phi}(r,R_{gcA}) - 2\Delta J^{\phi}(r,R_{gcAB})
+ \Delta J^{\phi}(r,R_{gcB}) \, ,
\end{split}
\end{gather}

\noindent where $2R_{gcAB}^2 = R_{gcA}^2 + R_{gcB}^2$.

Numerical calculations of the mesoscopic correlations at the level of
centers of mass show that $h^{cc}(r)$ is practically independent of
temperature.  This feature is consistent with the fact that
intermolecular interactions between centers of mass of two block
copolymers are unaffected by changes in concentration fluctuations,
given that monomer correlations arising from two blocks are averaged
out by the coarse-graining procedure.  Moreover, the effect is
intuitively explained by the fact that phase separation only occurs on
the microscopic scale of $R_g$. As a consequence,
$\Delta{h^{cc,\phi}}(r)\approx 0$.  In contrast, the interaction
between blocks, even for symmetric diblock copolymers, depends
strongly on changes with temperature and the proximity of the system
to the spinodal condition, as discussed in the previous sections.

\section{Numerical test of the coarse-graining procedure in the athermal limit}
\label{SX:COMP}
\noindent As a test of our coarse-graining expressions, we compare
theoretical predictions with computer simulation data \cite{JARAM} of
homopolymer melts in the athermal ($N\chi_{\mathit{eff}}= 0$) regime.
Specifically, we use trajectories of united atom molecular dynamics
(UA-MD) simulations of a polyethylene (PE) homopolymer melt composed
of chains with degree of polymerization $N=96$, total site number
density $\rho=0.0321\mbox{ \AA}^{-3}$, temperature $T=453\mbox{ K}$,
and $R_g = 16.78\mbox{ \AA}$ \cite{JARAM}.  Table \ref{TB:LENG} lists
the relevant length scales that enter into Eq.\ (\ref{EQ:LDEF}), which
are extracted from the simulation and used as an input to our
calculations.  By comparing theory against simulations in the high
temperature regime, where concentration fluctuations are not present,
we can test the ability that our description has in capturing the
effect of architectural asymmetry.  We consider a diblock copolymer
system where chain branches are of equal size (symmetric, $f=0.50$),
and where branches are of unequal size (asymmetric, $f=0.25$), testing
both intra- and intermolecular form factors at the block and
center-of-mass levels.

\begin{table}[htdp]
\centering
\caption[Length Scales for a Model Polyethylene Diblock
Copolymer]{Length Scales for a Model Polyethylene
Diblock Copolymer.}
\begin{tabular}{ccccc}
\hline \hline
\T \B Length [\AA] & $f = 0.50$  &  $f = 0.25$ \\
\hline
\T $R_{gAA}$   & 10.86  &  6.63 \\
   $R_{gAB}$   & 27.80  & 29.08 \\
   $R_{gBA}$   & 27.79  & 26.28 \\
\B $R_{gBB}$   & 10.88  & 14.12 \\
\hline \hline
\end{tabular}
\label{TB:LENG}
\end{table}

While form factors in a diblock copolymer liquid at the monomer level
have been extensively investigated, analytical expressions that
represent well the structure factors at the block level are not known.
As a starting point, we consider the block-monomer {\it
intra}molecular form factors, which are assumed in this paper to
follow Eq.\ (\ref{EQ:WBMK}). The latter is just a simple
implementation of the well-known approximation for the center-of-mass
monomer form factor in homopolymer melts \cite{YAMAKAWA}. In Fig.\
\ref{FG:WBMK}, we test the Gaussian form of $\omega^{bm}_{\alpha
\beta}(k)$ per Eq.\ (\ref{EQ:WBMK}) against simulation data for
domain-resolved contributions. The top panel in Fig.\ \ref{FG:WBMK}
displays data for a compositionally symmetric diblock copolymer, for
both self and cross contributions. The correlation between monomer and
block center-of-mass sites decays faster in cross contributions, where
the distance between the two species is larger than in the self
contribution. The middle and bottom panels display the same comparison
for data of a compositionally asymmetric diblock copolymer.  We
observe good agreement between the proposed expression, Eq.\
(\ref{EQ:WBMK}), and simulation data for both symmetric and asymmetric
diblocks.  The Gaussian shape of the curve holds for any block-monomer
form factor provided that the number of monomers in the block is
sufficiently high, and the central limit theorem applies. The Gaussian
form of the function allows for the analytical solution of the
intermolecular block-block and block-monomer form factors.  As a final
check, the test of the total contribution, Eq.\ (\ref{EQ:WBMT}),
against simulations shows that the Gaussian form of the total
intramolecular block-monomer structure factor also represents
simulation data well.

As a next step, we compare in Figs.\ \ref{FG:HBMaK} and
\ref{FG:HBMaR} our description of the {\it inter}molecular structure
factor at the block-monomer level with united atom molecular dynamics
simulation data. We find that the agreement of analytical expressions,
Eq.\ (\ref{EQ:hBMkup}) and Eqs.\ (\ref{EQ:Hbmdir}) to
(\ref{EQ:Hphitutti}), with simulations is satisfactory in both real
and reciprocal spaces.  For the asymmetric system, $f=0.25$, where the
$A$-block is comprised of only 24 monomeric sites, the agreement
between theory and simulation data tends to become rather qualitative
whenever a site in the $A$-block is involved, e.g.\ in
$h^{bm}_{AA}(k)$ and $h^{bm}_{AB}(k)$ the largest discrepancy is
localized near $kR_{g_{AA}}\sim2$ (Fig.\ \ref{FG:HBMaK}).  It is
interesting to note that while our approach predicts that
$h^{bm}_{AA}(k)$ and $h^{bm}_{AB}(k)$  are practically
indistinguishable in the compositionally asymmetric diblock, data are
sensitive to numerical errors due to finite-size effects.  This
discrepancy carries over to real space, where the data tends to be
underestimated by the corresponding functions (Fig.\ \ref{FG:HBMaR}).
However, the distinction between $h^{bm}_{AA}(r)$ and $h^{bm}_{AB}(r)$
is subtle, and our theoretical approach appears to provide a
reasonable description both in real and reciprocal space. Moreover,
the $h^{bm}_{BB}(r)$ and $h^{bm}_{BA}(r)$ terms are modeled rather
well since the $B$-block is not affected by finite-size effects, and
good agreement is also observed in reciprocal space.

Next, we consider a comparison against simulation data of the
theoretical block-block \textit{inter}molecular total correlation
functions.  In Fig.\ \ref{FG:HBBK}, we show the function in reciprocal
space, $h^{bb}_{\alpha\beta}(k)$, for both compositionally symmetric
($f=0.5$) and asymmetric ($f=0.25$) diblocks.  In both cases,
agreement is found, within numerical error, between Eq.\
(\ref{EQ:HBBAB}) and simulation data.  The top panel in the figure
depicts the analytical solution involving both the Pad\'e approximant
of the intramolecular structure factors, as well as its Debye
approximation, Eq.\ (\ref{EQ:DBYE}).  The Debye approximation appears
to give slightly better agreement with simulation than the Pad\'e
form. In turn, the Debye approximation yields only a minor improvement
for the real space representations, as shown in Fig.\ \ref{FG:HBBR}.
While the Debye approximation appears to model the data in slightly
better fashion than the Pad\'e approximant, the caveat in using it is
that the reciprocal space representation must be numerically inverted,
thereby defeating the purpose of obtaining an analytical solution.

Shown in Fig.\ \ref{FG:HBMtt} is the sum of contributions for the
block-monomer total pair correlation functions.  Upon inspection, it
becomes evident that finite-size effects of the $A$-block in the
asymmetric system cause a deviation from our theoretical predictions
for $r\sim 0.5\,R_{g_{AA}}$, which is consistent with our prior
findings for the domain-resolved functions.  For larger length scales,
however, the agreement is excellent.  Discrepancies due to finite-size
effects are not present in compositionally symmetric systems, which
are modeled overall rather well by our theory.

The sum represented by Eq.\ (\ref{EQ:HBTO}), which gives information
of the liquid as a whole at the level of molecular blocks, is slightly
sensitive to asymmetric features.  However, the theoretical
expressions are able to capture such small effects in both reciprocal
and real spaces, as shown in Fig.\ \ref{FG:HBBT}.  Specifically,
blocks of different size induce a break in symmetry in the liquid,
where packing is moderately favored at length scales smaller than the
overall spatial extension of the molecule.  While
$h^{mm}_{\mathit{tot}}(r)$ is identical for both symmetric and
asymmetric cases, effects emerge at the block level that depend on
differences in block size, a characteristic feature which enters in
our development by the behavior of intramolecular
$\omega^{bm}_{\alpha\beta}(k)$ form factors.

As a final test, we calculated the center-of-mass total pair
correlation function given by Eq.\ (\ref{EQ:HCCG}).  The agreement
between theory and simulations is good.  Shown also is the discrepancy
that arises when replacing $\omega^{cm}_{\mathit{tot}}(k)$ with 
Eq.\ (\ref{EQ:WCMT}) together with the Gaussian approximation of the
monomer-center-of-mass intramolecular structure factors, which results in a weak
underestimate of $h^{cc}(r)$, as indicated in Fig.\ \ref{FG:HCCR}.

\section{Temperature-dependent model calculations and clustering phenomena}
\label{SX:MODL}

\noindent In this Section we explore the
temperature effects which give rise to concentration fluctuations in
diblock copolymer systems.  Our main goal here is to develop a
qualitative understanding of the liquid behavior at the mesoscopic scale as the system evolves
toward its microphase separation transition. 
To make contact with the calculations performed in the athermal regime, and presented in the previous section, we focus in our model calculation on a "real" system, and we compute the cooling
curves for the polyethylene system studied in Section \ref{SX:COMP}. 

Input to our approach is the  static structure factor, $S^{\phi}(k)$, described at the monomer-level,
as a function of the order parameter $N \chi_{eff}$. As discussed in Section \ref{SX:THEO}, the static structure factor for a diblock copolymer 
liquid presents a peak that increases in intensity as the system approaches phase separation. At the 
temperature of the phase transition, only certain fluctuations become anomalously large, as the
liquid segregates on the molecular scale of the molecular radius-of-gyration, $k^* \approx R_g^{-1}$.
Leibler's mean-field approach predicts a second order phase transition characterized by the divergence 
of the peak in the static structure factor, 
$[S^{\phi}(k^*)]^{-1} \rightarrow 0$. However, for experimental systems, where polymer chains are finite, the second order phase transition is suppressed by the 
onset of concentration fluctuation stabilization, and a first order phase transition is observed over the entire composition range.

At the monomer level, concentration fluctuation stabilization corrections are well accounted for by Brazovskii's correction to Landau approach, as described by Fredrickson and Helfand, as well as by the
integral equation theory PRISM developed by Schweizer, Curro and coworkers. Both theories predict the same scaling with $N$ of the static structure factor, namely   at high temperature ($T>> T_{ODT}$), random mixing applies and $S^{\phi}(k^*) \propto N$, while at the transition ($T\approx T_{ODT}$) they predict $S^{\phi}(k^*)/N \propto N^{1/3}$.
The mean-field behavior is recovered in the limit of infinite chain length. 
The choice of the model used as an input, at the monomer level, is not crucial, however to preserve the consistency of our formalism, we
correspondingly compute the cooling curve for the coarse-grained diblock in the framework of PRISM integral equation approach. The peak scattering intensity changes with temperature following the equation\cite{MDCMP,Macromol},
\begin{equation}
\frac{t}{(N_{\mathit{eff}})^{1/2}}s^{1/2}(k^*)
\left[s^{1/2}(k^*)-1\right] +
\left[1-t\right]s(k^*) - 1 = 0\, ,
\label{EQ:SNRM}
\end{equation}

\noindent where the form factor is
normalized by its athermal value, $s(k^*)=S^{\phi}(k^*)/S^{\phi}_0(k^*)$, and the
temperature, $t=T_{hta,s}/T$, is rescaled with respect to the spinodal
temperature obtained by applying the athermal, molecular PY closure.\cite{eddavid} 
Given the general relation $N \chi \propto T^{-1}$, the temperature $T_{hta,s}$ is calculated in the framework of PRISM theory from the ratio
$\chi_s/\chi_{hta,s} = (1+\xi_{\rho}/a)^{-1}$, 
Here $a$ is the spatial range of the tail in the Yukawa potential governing $AB$ interactions as  $v_{AB}(r) = (a/r)\varepsilon e^{-r/a}$, with
$\varepsilon>0$ being the interaction strength.  
For our calculations of polyethylene-like diblock, we set $a=0.5\sigma$, since this choice mimics the spatial range of the
Lennard-Jones potential\cite{MDCMP,Macromol}, and $\xi_{\rho}=0.346\sigma$.

The other term entering Eq.\ (\ref{EQ:SNRM}) that needs to be specified is
\begin{equation}
N_{\mathit{eff}} =
\frac{\overline{N}}{[\mathit{\Theta}(1+\xi_{\rho}/a)]^2}\, .
\label{EQ:NEFF}
\end{equation}

\noindent where $\overline{N} = N(\rho\sigma^3)^2$ is the Ginzburg
parameter that controls the strength of the concentration stabilization effect in
$S^{\phi}(k^*)$.  In addition, the
parameter
\begin{equation}
\mathit{\Theta} = \frac{x^*}{cf(1-f)(2N\chi_s)^{1/2}}\, ,
\end{equation}

\noindent depends on $c^2 = x^*\partial^2 F(x)/\partial x^2|_{x=x^*}$
which is of $\mathcal{O}(1)$.  Using the tabulated values from Ref.\
\cite{FRDIK}, $c = 1.102,1.278$ and $N\chi_s=10.495,18.122$ for
$f=0.50,0.25$, respectively.
The calculation described so far is standard in PRISM theory and examples have been reported in several papers.\cite{eddavid, eddavid1,MDCMP,Macromol} 

The cooling curves, obtained from the self-consistent solution of Eq.(\ref{EQ:SNRM}) for the system in this study, 
are presented in Fig.\ \ref{FG:CURV}.  
Note how concentration effects stabilize the response
in $S^{\phi}(k^*)^{-1}$ for the finite-size PE system investigated, as evidenced
by a leveling off upon further cooling.  There is a subtle
difference between the behavior for $S^{\phi}(k^*)$ for the $f=0.25$ and
$f=0.50$ cases, i.e.\ the two curves are indiscernible given the resolution in the figure. Also reported is the
mean-field prediction, which shows divergence of the structure factor at the spinodal temperature.

The system investigated is a diblock copolymer with fixed total number of monomers, $N=96$, identical segment 
lengths for the two blocks, $\sigma_A=\sigma_B$, and a repulsive Yukawa interaction between unlike monomers.
The chain is partitioned, first as a compositionally \textit{symmetric} diblock ($f=0.5$ and $N_A=N_B=48$) and then as a
compositionally \textit{asymmetric} diblock ($f=0.25$ and $N_A=24$ and $N_B=72$). Input to our coarse-graining theory
are the values of $S^{\phi}(k^*)$ calculated for these two systems at
$N\chi_{\mathit{eff}}/N\chi_{s} \in \{0.0, 0.5, 1.0, 2.0\}$, as indicated in
Fig.\ \ref{FG:CURV}. These values sample our systems in a range of temperatures that include
the athermal limit, $N\chi_{\mathit{eff}}/N\chi_{s} =0$, the spinodal temperature, $N\chi_{\mathit{eff}}/N\chi_{s}=1$, and the weak segregation limit down to (roughly) the ODT temperature, $N\chi_{\mathit{eff}}/N\chi_{s}=2$, calculated following the procedure
in Refs. \cite{MDCMP,Macromol}.

To study the effects of concentration fluctuations at the level of
blocks, we focus on the physical clustering of like species as defined in Eq.\ (\ref{deltagbb}).
In Fig.\ \ref{FG:HTMP}, $\Delta
g^{bb}(r)$ is shown for the symmetric and asymmetric cases.  At
athermal conditions concentration fluctuation contributions are absent. Repulsive
interactions between unlike monomers are screened and entropic contributions to the free energy
are dominant. The symmetric case  exhibits no local clustering
effects, and the system packs in an entirely random fashion.  For the  model diblock copolymer investigated in this study, where monomer 
bond lengths for the two blocks are equal ($\sigma_{A}=\sigma_B$), the two blocks
at high temperature are identical for a compositionally symmetric diblock, i.e.  $\Delta g^{bb}(r)=0$ 
for $f=0.5$.

For the
asymmetric case, on the other hand, there is an emergence of entropic
packing effects arising from the difference in block sizes, yielding a
response in $\Delta g^{bb}(r)$ where $AB$ contacts are favored ( $\Delta g^{bb}(0) < 0$ at high temperature for $f \neq 0.5$).

This
effect is quite apparent in the block frame of reference, which is
sensitive to local microdomain scale correlations.  Upon decreasing temperature,
the formation of self contacts, $AA$ and $BB$, becomes
energetically favorable as the system approaches its microphase
segregation transition ($\Delta g^{bb}(0) > 0$).

A shallow minimum develops with decreasing temperature, at a length scale corresponding to the size
of the microdomain, $r \approx 1.5 R_g$ for the symmetric case, because at the 
interface of the two domains
the number of  contacts between unlike species is higher than the number of self-contacts, i.e.\
$\Delta g^{bb}(r)< 0$. For compositionally asymmetric diblock copolymers, physical clustering occurs
around the minority species, and the minimum is slightly shifted toward the small-$r$ region.
In both cases, the minimum is smooth and shallow, indicating that there is no a sharp transition at the interface between $A$ and $B$ domains, which is a characteristic feature of the weak 
segregation regime: fluctuations still partially disorder the liquid, while it becomes increasingly correlated approaching its phase transition.

\section{Conclusion}
\noindent We have presented an analytical coarse-grained description
for diblock copolymer melts, where the blocks in a polymer molecule
are envisioned as two soft colloidal spheres connected by an entropic
spring.  Corresponding domain-resolved intermolecular total pair
correlation functions are formally derived from an integral equation
approach by solving the Ornstein-Zernike equation, which is formalized
as a matrix of monomer and block center-of-mass sites.  The total
pair correlation function describing the interactions occurring at the
center of mass of the molecule is also presented.

Analytical expressions for the total correlation functions of the
coarse-grained diblock  for a copolymer chain represented as a
Gaussian thread of vanishing thickness, with an interaction symmetric
potential acting between blocks of like and unlike chemical species,
are derived.  In the framework of this model, the analytical total
correlation functions contain contributions from density and
concentration fluctuations.  The concentration fluctuation terms
increase in intensity as the diblock melt approaches its microscopic
separation transition, however these do not diverge since finite-size
fluctuations suppress the second-order phase transition. The
contribution from concentration fluctuations drives the isotropic
clustering of like species as the system approaches its phase
transition.  In the athermal regime, where density fluctuations are
dominant, asymmetry in block size induces partial clustering of the
domains.

As a test of the coarse-graining approach, analytical expressions are
compared with data obtained from a symmetric diblock copolymer melt in
athermal conditions.  Our theoretical study shows that good agreement
is attained in real and reciprocal spaces for the total intermolecular
pair correlation functions at the block and center-of-mass levels.
Comparisons are made for a diblock copolymer melt composed of chains
with equally-sized, as well as unequally-sized, chain branches.

The present development corresponds to a significant stride in
presenting an analytical coarse-graining scheme for diblock copolymer
melts.  From our previous work, which has mainly focused on the
mesoscopic treatment of polymers at the center-of-mass level, the
results reported herein offer an intermediate level of coarse-graining
that preserves some detailed intramolecular information to account for
the physics proper of block copolymers.

\section{Acknowledgements}
\noindent United atom molecular dynamics simulation trajectories are a
courtesy of G.\ S.\ Grest, J. G. Curro, and E.\ Jaramillo from Sandia
National Laboratories.  The authors are grateful to the National
Science Foundation (NSF) under Grant No.\ DMR-0207949 for financial
support.  Acknowledgement is made to the Donors of the American
Chemical Society Petroleum Research Fund for partial support of this
research.  In addition, E.\ J.\ S.\ is indebted to the NSF for a
Graduate Research Fellowship.

\appendix

\section{Auxiliary Functions for Real-Space Representations}
\label{AP:AFXN}
\subsection{Density Fluctuation Terms}
\label{AP:DENS}
\noindent We collect here representations for the auxiliary functions
used in the main text.  The density fluctuation contribution arising
from coupled frames of reference (i.e.\ those arising between a
mesoscopic level and the local monomer level) are represented by
\begin{gather}
\begin{split}
I^{\rho}(r,R) & = -\frac{\xi_{\rho}'}{2r}
\left(1-\xi_c^2/\xi_{\rho}^2\right)
e^{R^2/(6\xi_{\rho}^2)} \\
& \times
\Bigg[
e^{r/\xi_{\rho}}\mbox{erfc}
\left(\frac{R}{\sqrt{6}\,\xi_{\rho}}
+\frac{\sqrt{3}\,r}{\sqrt{2}\, R}\right)
-e^{-r/\xi_{\rho}}\mbox{erfc}
\left(\frac{R}{\sqrt{6}\,\xi_{\rho}}
-\frac{\sqrt{3}\,r}{\sqrt{2}\,R}\right)
\Bigg] \, .
\label{EQ:Hrho}
\end{split}
\end{gather}

\noindent Density fluctuations of self-character (i.e.\ contributions
arising between the same mesoscopic level), are given by the relation
\begin{eqnarray}
J^{\rho}(r, R) & = &
\frac{3}{2}\sqrt{\frac{3}{\pi}}
\frac{\xi_{\rho}'}{R}
\left(\frac{\xi_c}{R}\right)^2
\left(1-\frac{\xi_c^2}{\xi_{\rho}^2}\right)
e^{-3r^2/(4R^2)}
-\frac{\xi_{\rho}'}{2r}
\left(1-\xi_c^2/\xi_{\rho}^2\right)
e^{R^2/(3\xi_{\rho}^2)} \nonumber\\
& \times&
\Bigg[
e^{r/\xi_{\rho}}\mbox{erfc}
\left(\frac{R}{\sqrt{3}\,\xi_{\rho}}
+\frac{\sqrt{3}\,r}{{2}\, R}\right)
-e^{-r/\xi_{\rho}}\mbox{erfc}
\left(\frac{R}{\sqrt{3}\,\xi_{\rho}}
-\frac{\sqrt{3}\,r}{{2}\,R}\right)
\Bigg] \, .
\label{EQ:GEN}
\end{eqnarray}

\subsection{Concentration Fluctuation Terms}
\label{AP:CONC}
\noindent In the section, we summarize analytic expressions
representing the contribution from concentration fluctuations.
Following the previous arrangement, fluctuations of mixed character
are described by the expression
\begin{gather}
\begin{split}
I^{\phi}(r,R) & =
3 \sqrt{\frac{6}{\pi}}f(1-f)
\frac{\xi_{\rho}'}{R}
\left(\frac{\xi_c}{R}\right)^2
e^{-3r^2/(2R^2)} +I'^{\phi}(r,k_{+},R) - I'^{\phi} (r,k_{-},R)
 \label{EQ:gener}
\end{split}
\end{gather}

\noindent with
\begin{gather}
\begin{split}
I'^{\phi} (r,k_{\pm},R) & = {\mp} f(1-f)\frac{\xi_{\rho}'}{2r}
\left(1-\xi_c^2k_{\pm}^2\right)
\left(\frac{k_{\pm}^2}{k_+^2-k_-^2}\right)
e^{R^2k_{\pm}^2/6} \\
& \times \Bigg[
e^{rk_{\pm}}\mbox{erfc}
\left(\frac{Rk_{\pm}}{\sqrt{6}}
+\frac{\sqrt{3}\,r}{\sqrt{2}\, R}\right)
-e^{-rk_{\pm}}\mbox{erfc}
\left(\frac{Rk_{\pm}}{\sqrt{6}}
-\frac{\sqrt{3}\,r}{\sqrt{2}\,R}\right)
\Bigg] \, .
\end{split}
\end{gather}

\noindent Analogously, the concentration fluctuation contribution of self
character is represented by the relations
\begin{multline}
J^{\phi}(r, R) =
f(1-f) \frac{3}{2}  \sqrt{\frac{3}{\pi}}
\frac{\xi_{\rho}'}{R}
\left(\frac{\xi_c}{R} \right)^2 e^{-3r^2/(4R^2)} \\
 \left[ 2 +   \frac{9\xi_c^2}{2R^2}
\left( 1- \frac{r^2}{2R^2} \right)
- f(1-f) \frac{ N}{S(k^*)} + \sqrt{\frac{3}{f(1-f)}}\,\,\right]
+ J'^{\phi}(r,k_{+},R) -J'^{\phi}(r,k_{-},R) \, ,
\label{EQ:jei}
\end{multline}

\noindent with
\begin{gather}
\begin{split}
J'^{\phi}(r,k_{\pm},R) & ={\mp} f(1-f)\frac{\xi_{\rho}'}{2r}
\left(1-\xi_c^2k_{\pm}^2\right)^2
\left(\frac{k_{\pm}^2}{k_+^2-k_-^2}\right)
e^{R^2k_{\pm}^2/3} \\
&\times \Bigg[
e^{rk_{\pm}}\mbox{erfc}
\left(\frac{Rk_{\pm}}{\sqrt{3}}
+\frac{\sqrt{3}\,r}{2R}\right)
-e^{-rk_{\pm}}\mbox{erfc}
\left(\frac{Rk_{\pm}}{\sqrt{3}}
-\frac{\sqrt{3}\,r}{2R}\right)
\Bigg] \, ,
\end{split}
\end{gather}

\noindent with
\begin{eqnarray}
k_{\pm} = \xi_1^{-1}\mp i(\xi_2)^{-1}.
\label{eqcsi}
\end{eqnarray}

\noindent 
In these expressions the temperature
dependence enters through $ N/S(k^*)$ and the values of $k_{\pm}$
evaluated at the reference athermal and thermal states.
The numerical evaluation of the
complementary error function with complex arguments  is more generally known 
in the context of Faddeeva's function
in the field of optics, and poses no problem.  However we note that, 
while the length scale associated with
$k_{\pm}$ is complex, it is only a consequence of the factorization
given by Eq.\ (\ref{EQ:SKPHI}). In our expressions the imaginary 
components strictly vanish when taking into consideration the positive and
negative branches of the functions.

\section{Simplified Formalism to Coarse Grain Compositionally
Symmetric Diblock Copolymers}
\label{AP:SYMM}
\noindent For a compositionally symmetric diblock copolymer ($f=0.5$),
the general coarse-graining formalism presented in the main text
becomes quite simple, since the equalities
$\omega^{mm}_{AA}(k)=\omega^{mm}_{BB}(k)$,
$\omega^{bm}_{AB}(k)=\omega^{bm}_{BA}(k)$, and
$\omega^{bm}_{AA}(k)=\omega^{bm}_{BB}(k)$ apply.  Also, we have that
$h^{mm}_{AA}(k)=h^{mm}_{BB}(k)$, as well as
$h^{bb}_{AA}(k)=h^{bb}_{BB}(k)$. By enforcing these rules, we obtain
\begin{align}
h^{bm}_{AA}(k)
& = \left[\frac{\omega^{bm}_{\mathit{tot}}(k)}
{\omega^{mm}_{\mathit{tot}}(k)}\right]
h^{\rho}(k)
+ 2 \left[\frac{\omega^{bm}_{AA}(k)-\omega^{bm}_{AB}(k)}
{\omega^{mm}_{\mathit{tot}}(k)}\right]
\Delta h^{\phi}(k)\  ,
\nonumber \\ 
h^{bm}_{AB}(k)
& = \left[\frac{\omega^{bm}_{\mathit{tot}}(k)}
{\omega^{mm}_{\mathit{tot}}(k)}\right]
h^{\rho}(k)
- 2\left[\frac{\omega^{bm}_{AA}(k)-\omega^{bm}_{AB}(k)}
{\omega^{mm}_{\mathit{tot}}(k)}\right]
\Delta h^{\phi}(k)\ ,
\label{EQ:HBMBS}
\end{align}

\noindent and
\begin{align}
h^{bb}_{AA}(k)
& = \left[\frac{\omega^{bm}_{\mathit{tot}}(k)}
{\omega^{mm}_{\mathit{tot}}(k)}\right]^2
h^{\rho}(k)
+ 4 \left[\frac{\omega^{bm}_{AA}(k)-\omega^{bm}_{AB}(k)}
{\omega^{mm}_{\mathit{tot}}(k)}\right]^2
\Delta h^{\phi}(k)\, ,
\nonumber \\
h^{bb}_{AB}(k)
& = \left[\frac{\omega^{bm}_{\mathit{tot}}(k)}
{\omega^{mm}_{\mathit{tot}}(k)}\right]^2
h^{\rho}(k)
- 4\left[\frac{\omega^{bm}_{AA}(k)-\omega^{bm}_{AB}(k)}
{\omega^{mm}_{\mathit{tot}}(k)}\right]^2
\Delta h^{\phi}(k)
\, .
\label{EQ:HBBS}
\end{align}

\noindent In real space, the block-monomer and block-block
intermolecular total pair correlation functions  separate into density
and concentration fluctuation contributions.  For the block-monomer
contributions, density and concentration fluctuations become,
respectively,
\begin{align}
{h^{bm,\rho}_{\alpha \alpha}}(r) & =
\frac{1}{2}I^{\rho}(r,R_{g_{\alpha A}}) +
\frac{1}{2}I^{\rho}(r,R_{g_{\alpha B}})
= {h^{bm,\rho}_{\alpha \beta}}(r) \, ,
\end{align}

\noindent and
\begin{align}
{\Delta h^{bm,\phi}_{AA}}(r) & =
{\Delta h^{bm,\phi}_{BB}}(r)=
  2 \Delta I^{\phi}(r,R_{gAA})
 - 2\Delta I^{\phi}(r,R_{gAB}) \nonumber \\
 & =  -{\Delta h^{bm,\phi}_{AB}}(r) =
  -{\Delta h^{bm,\phi}_{BA}}(r)\ .
\end{align}

\noindent Moreover, the density fluctuation contribution for the
block-block correlation function is given by
\begin{align}
{h^{bb,\rho}_{AA}}(r) &= \frac{1}{4}J^{\rho}(r,R_{AAAA}) +
\frac{1}{2}J^{\rho}(r,R_{AAAB}) + \frac{1}{4}J^{\rho}(r,R_{ABAB})
= {h^{bb,\rho}_{AB}}(r)\, ,
\end{align}

\noindent while the corresponding concentration fluctuation
contribution is
\begin{align}
\Delta{h^{bb,\phi}_{AA}}(r) &=
\Delta{J}^{\phi}(r,R_{AAAA})
-2\Delta{J}^{\phi}(r,R_{AAAB})
+\Delta{J}^{\phi}(r,R_{ABAB}) = -\Delta{h^{bb,\phi}_{AB}}(r)\ .
\end{align}

\noindent The functions $I^{\rho}(r,R_{g\alpha \beta})$,
${J}^{\rho}(r,R_{\alpha \beta \gamma \delta})  $,
$I^{\phi}(r,R_{g\alpha \beta})$, and ${J}^{\phi}(r,R_{\alpha \beta
\gamma \delta})$ are defined in Section \ref{AP:AFXN} of the Appendix,
with $\alpha, \beta, \gamma, \delta \in \{A,B\}$.

\newpage
FIGURE CAPTIONS:\\

\begin{figure}[]
\centering
\includegraphics[scale=0.275,angle=-90]{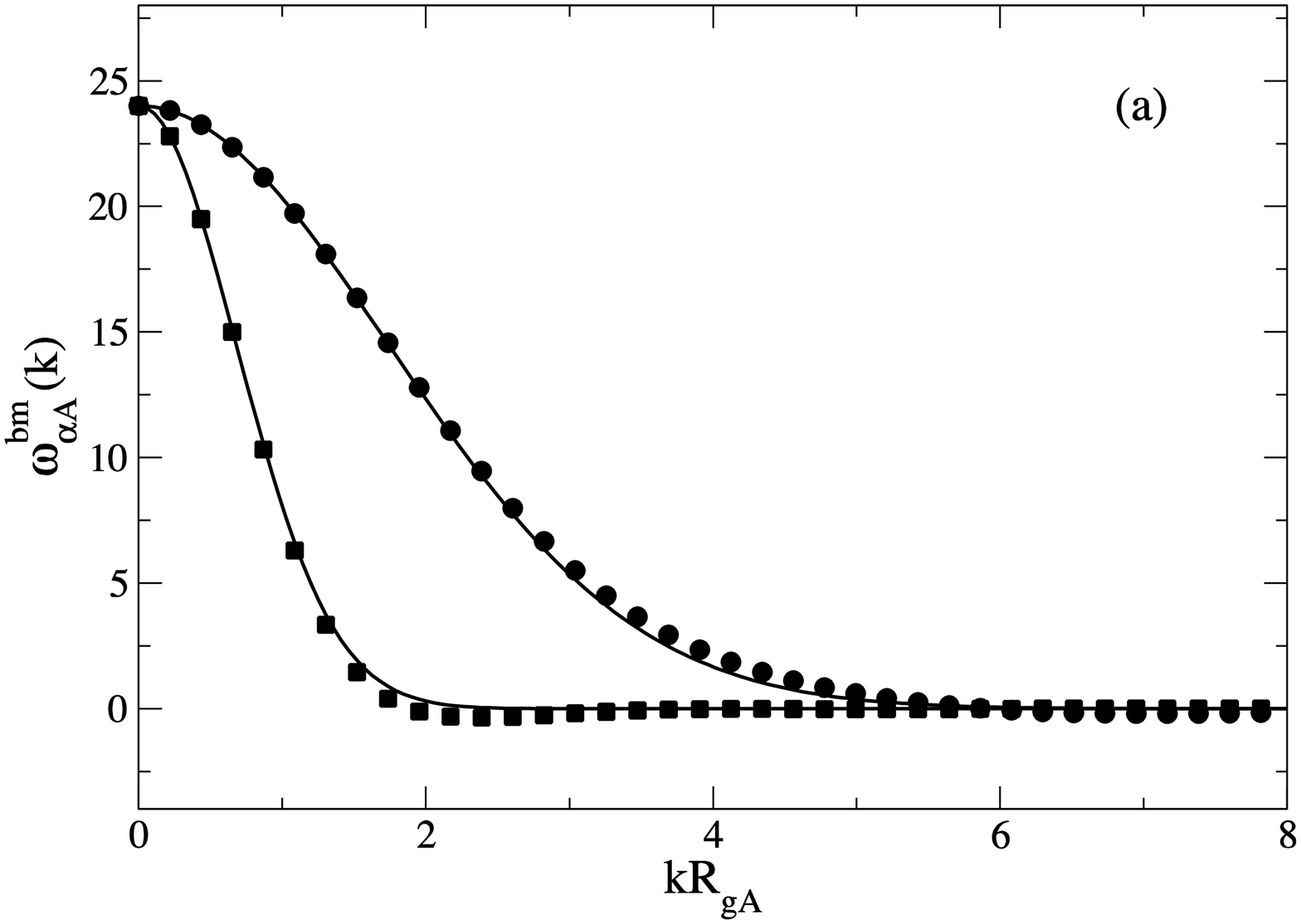}
\\
\includegraphics[scale=0.275,angle=-90]{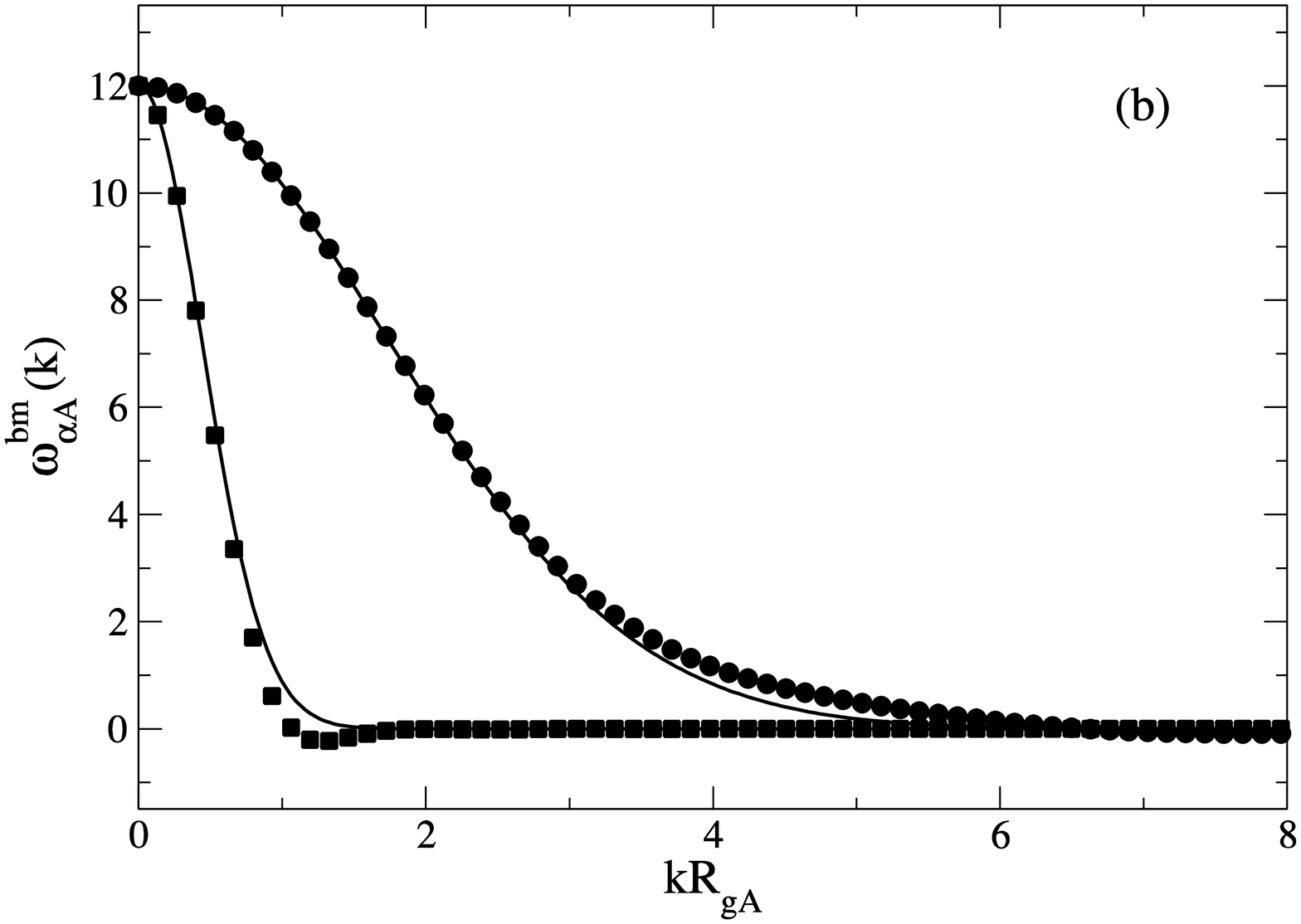}
\\
\includegraphics[scale=0.275,angle=-90]{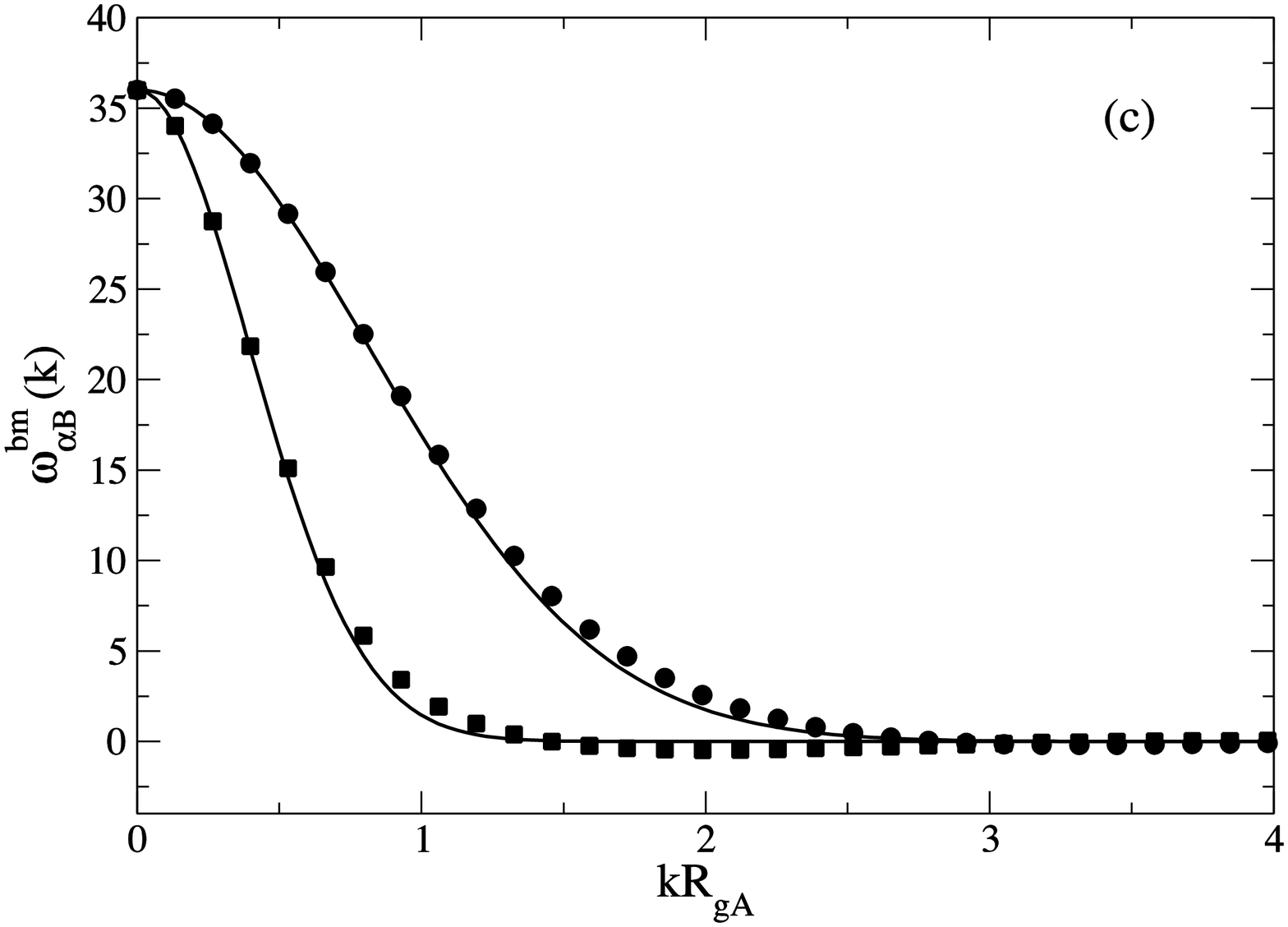}
\caption{}
\label{FG:WBMK}
\end{figure}

FIG.\ \ref{FG:WBMK}
Plot of $\omega^{bm}_{\alpha\beta}(k)$.  Shown are the Gaussian
representations [lines] from theory and data from united atom
molecular dynamics simulations correspondingly.  The $f=0.50$ case is
shown in Panel (a), whereas the $f=0.25$ case is shown in Panels (b)
and (c).  Data is resolved into self- [circles] and
cross-contributions [squares]. Panels (a) and (b), both display the
self, $\omega^{bm}_{AA}(k)$, and the cross,  $\omega^{bm}_{BA}(k)$,
block-monomer contributions. In Panel (c), $\omega^{bm}_{BB}(k)$ is
the self and $\omega^{bm}_{AB}(k)$ is the cross contribution.\\

\begin{figure}
\centering
\includegraphics[scale=0.4,angle=-90]{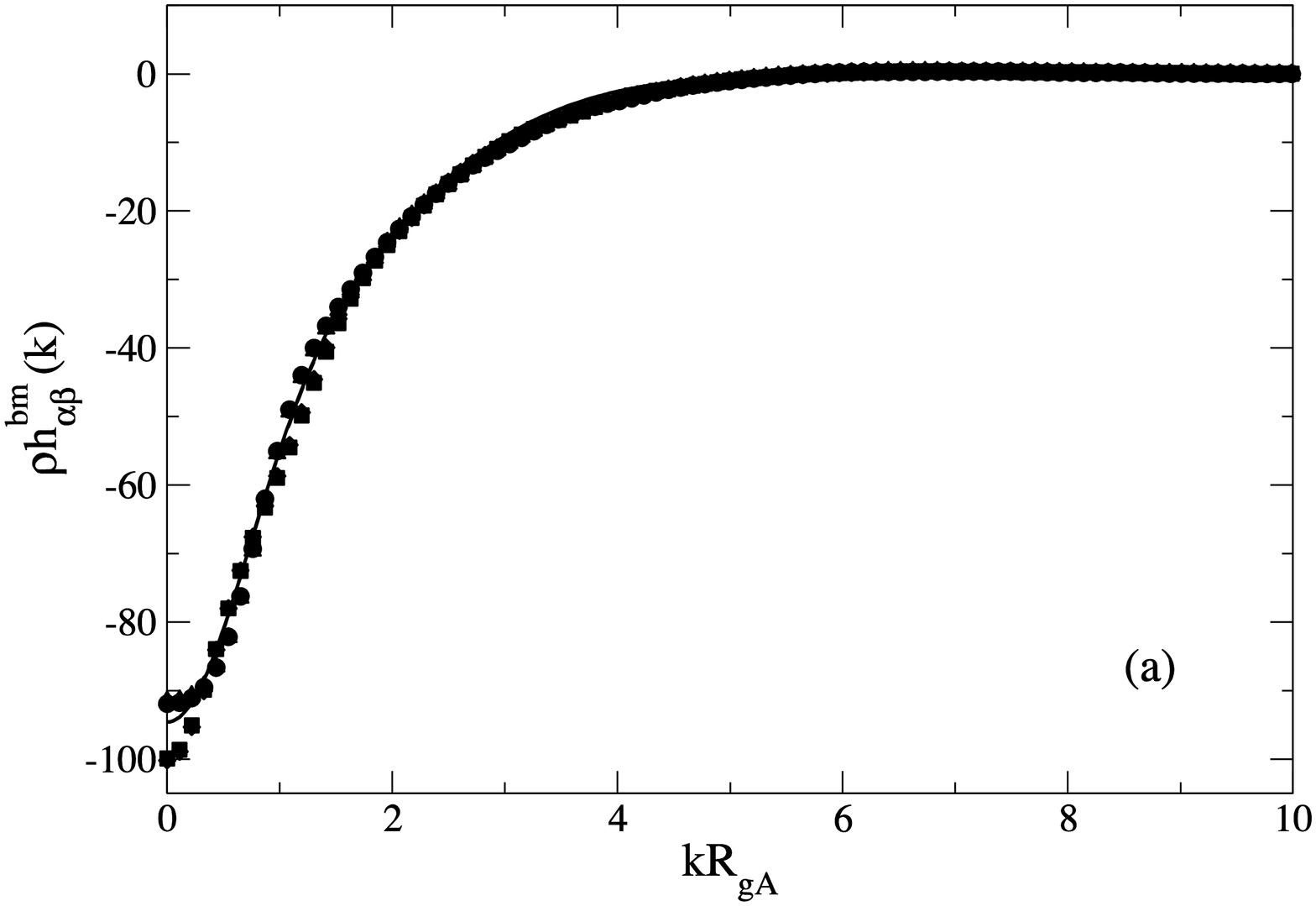}
\\
\includegraphics[scale=0.4,angle=-90]{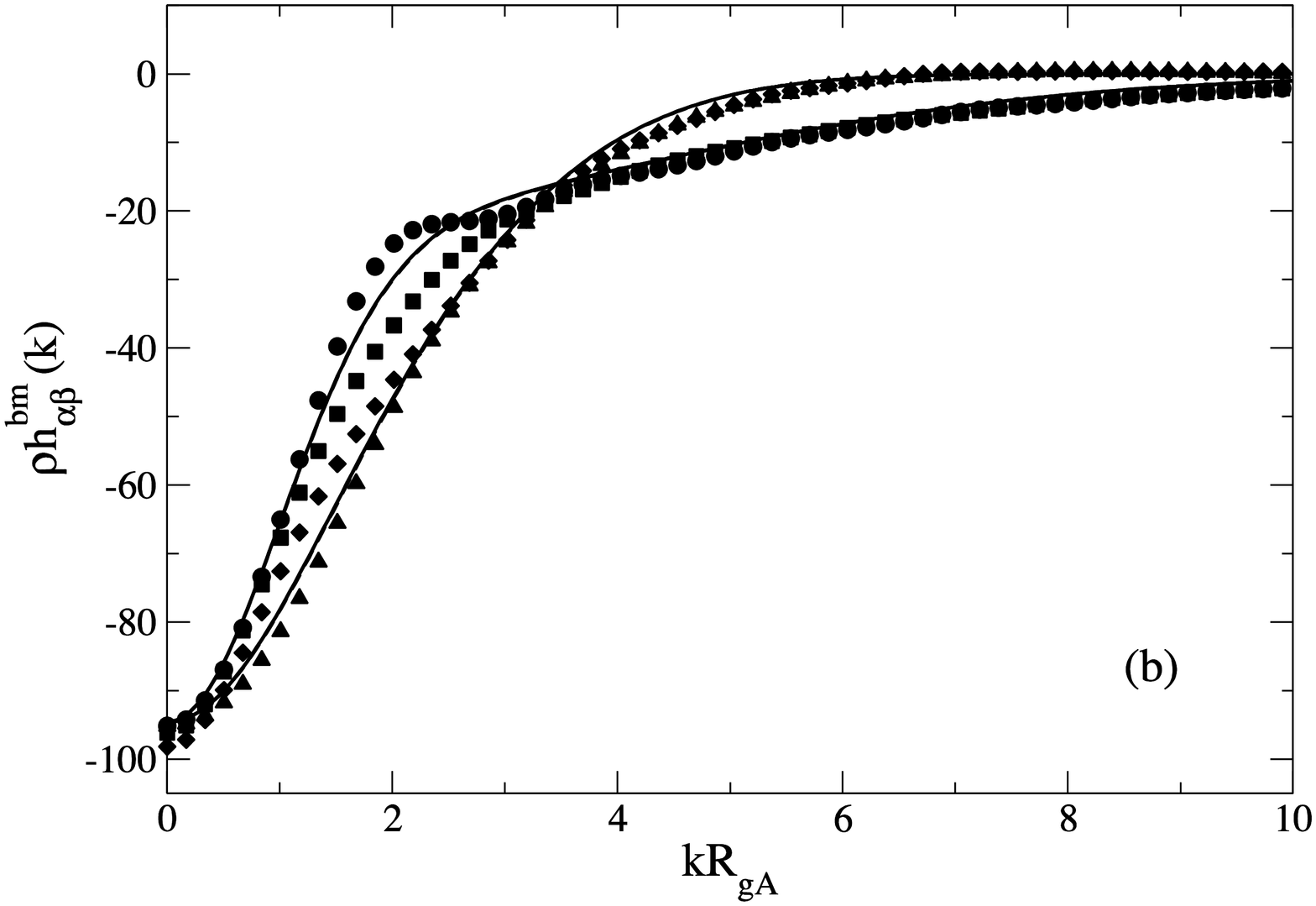}
\caption{}
\label{FG:HBMaK}
\end{figure}

FIG.\ \ref{FG:HBMaK}
Plot of $h^{bm}_{\alpha\beta}(k)$.  Shown are theoretical
representations [lines] along with data from united atom molecular
dynamics simulation [symbols]: $AA$ [circles], $AB$ [squares], $BA$
[diamonds], and $BB$ [triangles] contributions.  Panel (a) is for
$f=0.50$, while Panel (b) is for $f=0.25$.  The dashed lines (which
are indistinguishable in the plots) correspond to the solutions
obtained from the Debye representation of
$\omega^{mm}_{\alpha\beta}(k)$.\\

\begin{figure}[]
\centering
\includegraphics[scale=0.4,angle=-90]{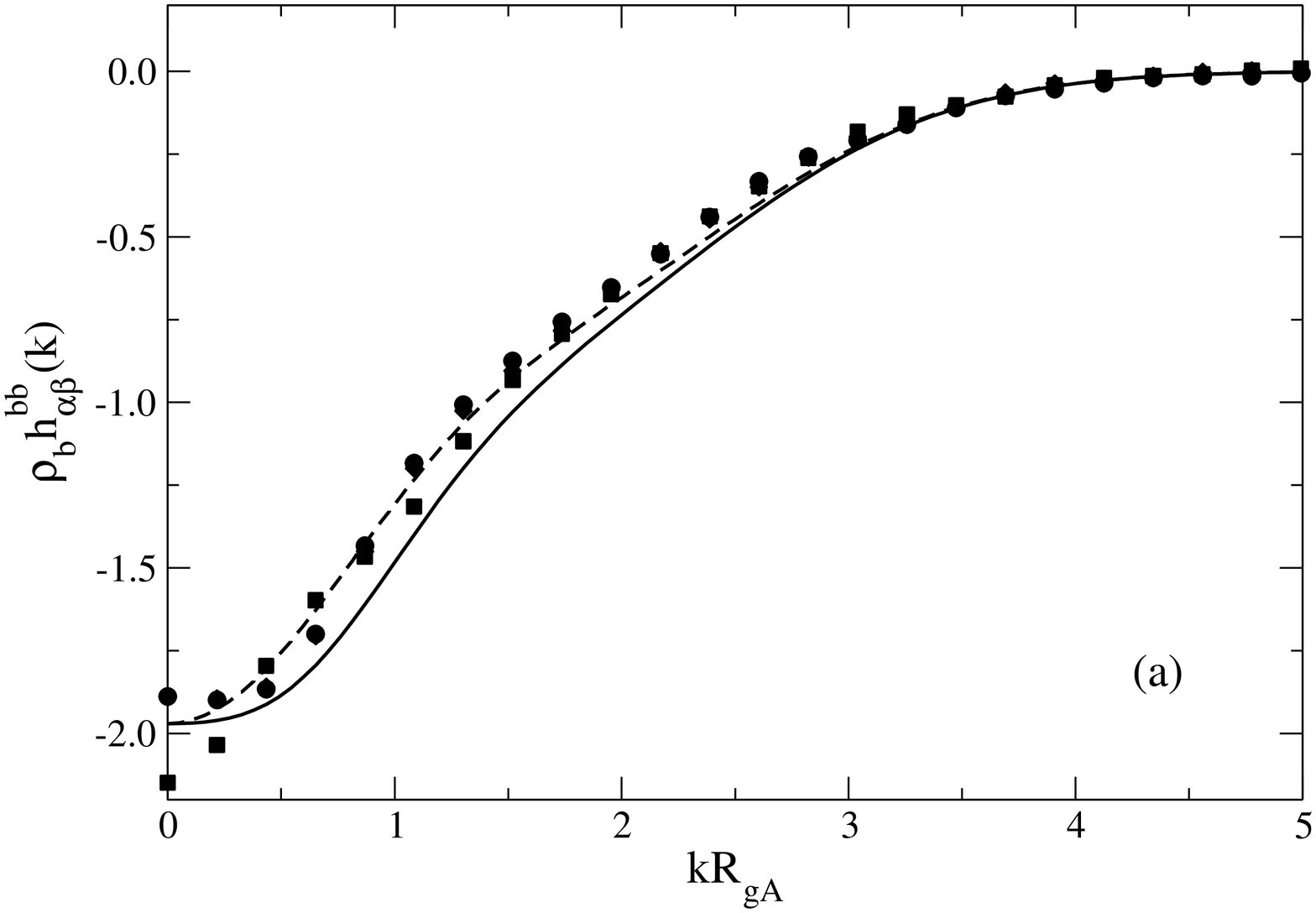}
\\
\includegraphics[scale=0.4,angle=-90]{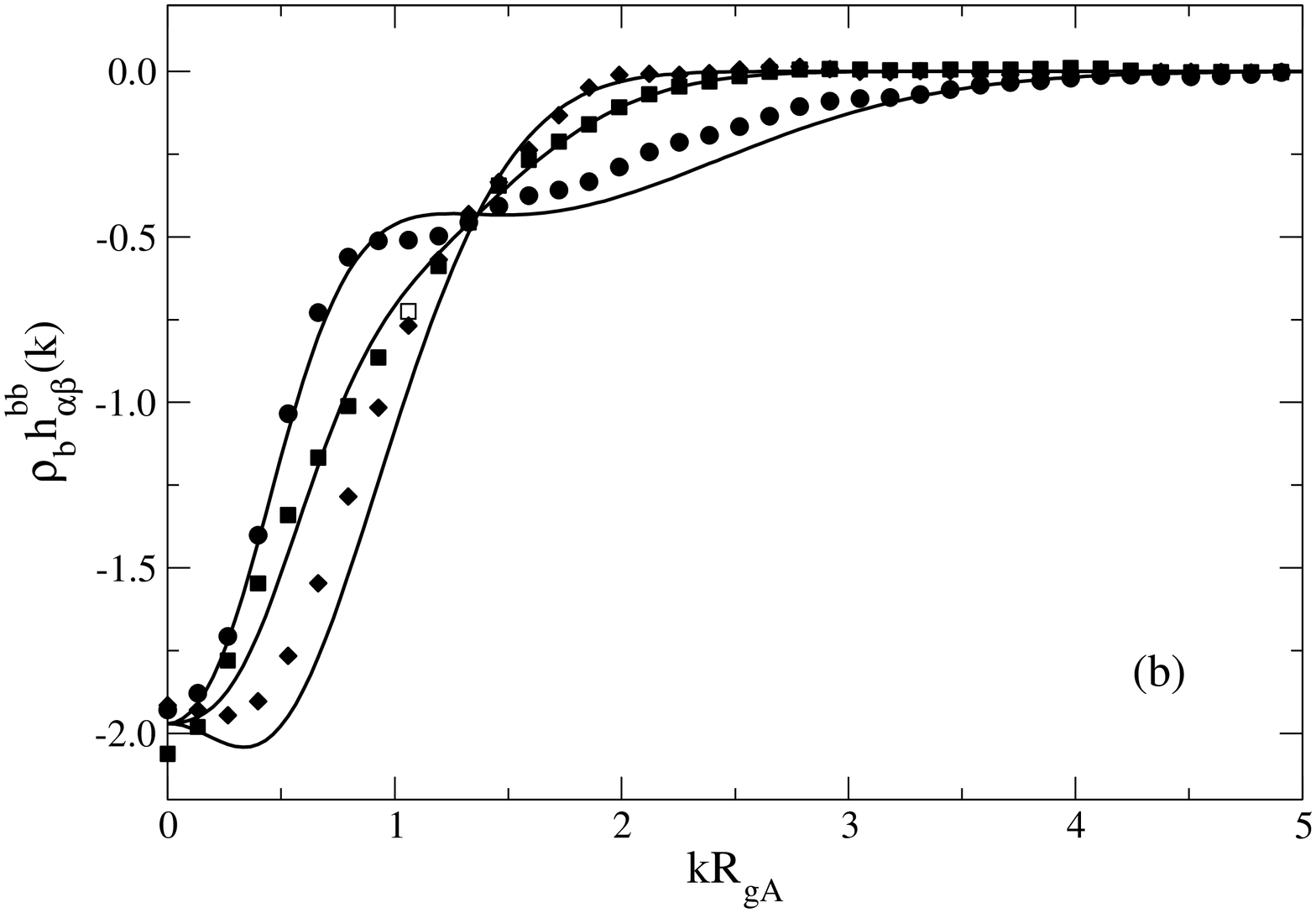}
\caption{}
\label{FG:HBBK}
\end{figure}

FIG.\ \ref{FG:HBBK}
Plot of $h^{bb}_{\alpha\beta}(k)$.  Shown are the theoretical
representations [lines] along with data from united atom molecular
dynamics simulations [symbols]: $AA$ [circles], $AB$ [squares], and
$BB$ [diamonds] contributions.  Panel (a) is for $f=0.50$, while Panel
(b) is for $f=0.25$.  Panel (a) also shows the result from the Debye
representation of $\omega^{mm}_{\alpha\beta}(k)$ [dashed line].\\

\begin{figure}
\centering
\includegraphics[scale=0.4,angle=-90]{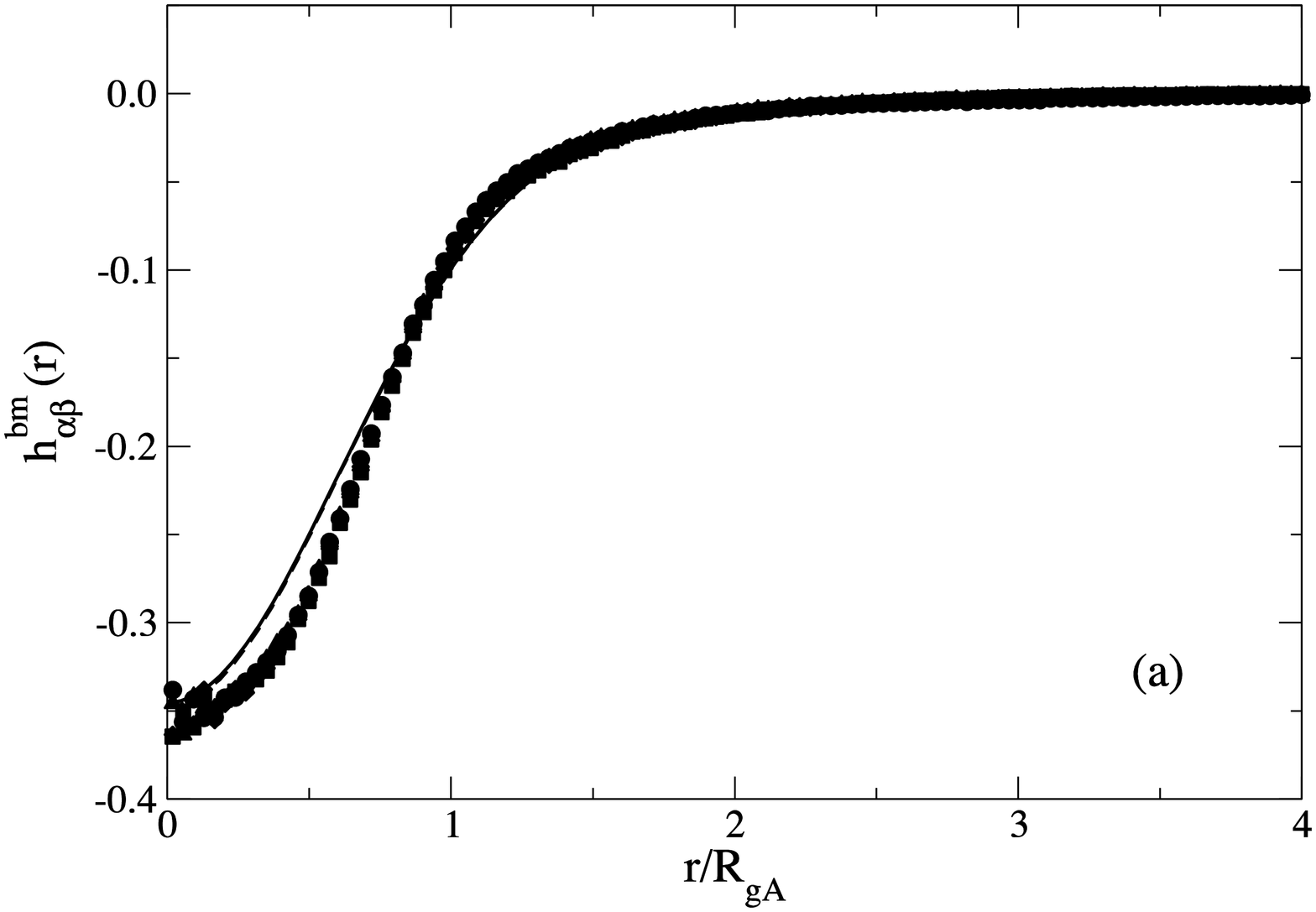}
\\
\includegraphics[scale=0.4,angle=-90]{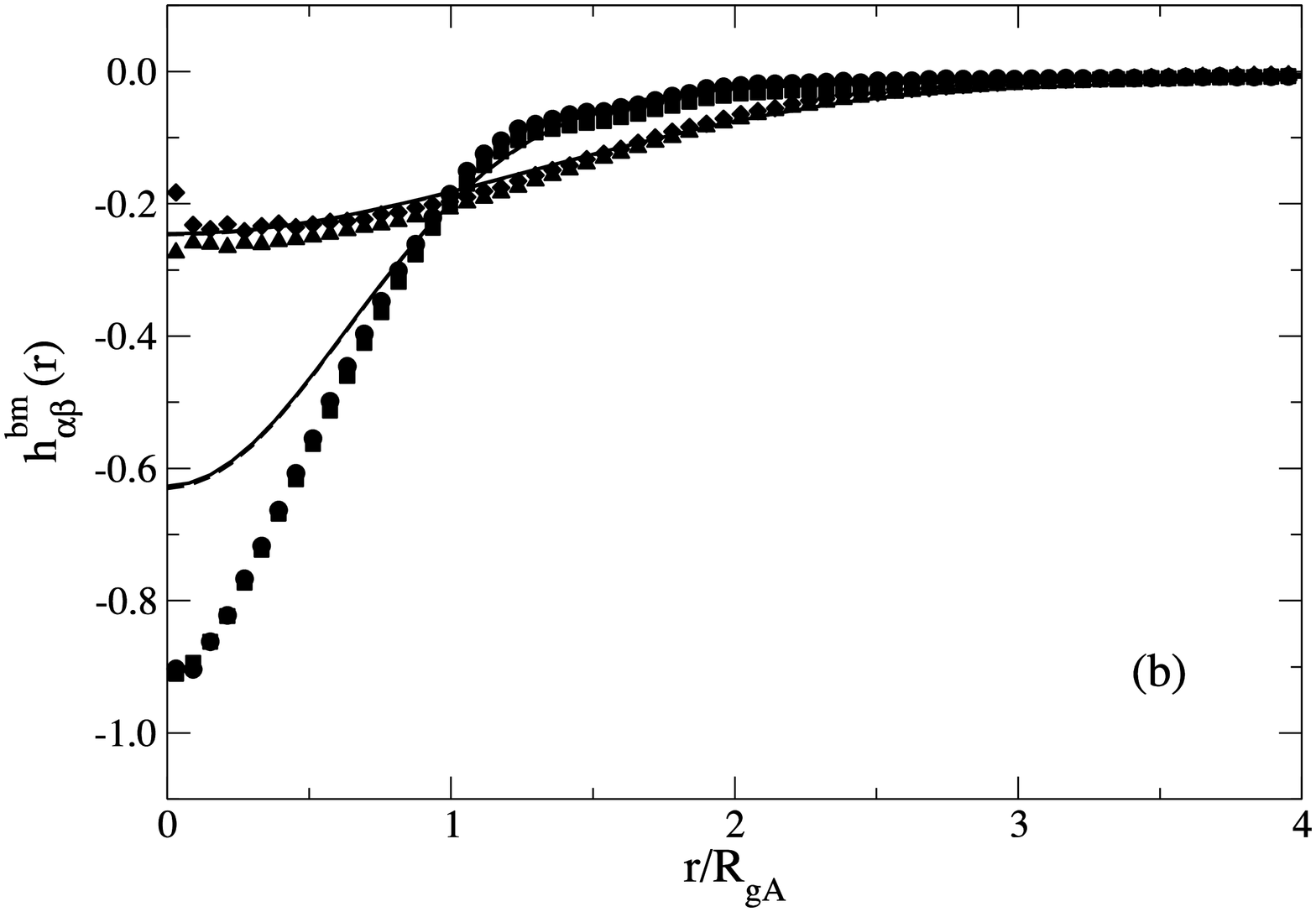}
\caption{}
\label{FG:HBMaR}
\end{figure}

FIG.\ \ref{FG:HBMaR}
Plot of $h^{bm}_{\alpha\beta}(r)$.  Data is arranged as in Fig.\
\ref{FG:HBMaK}.\\

\begin{figure}[]
\centering
\includegraphics[scale=0.4,angle=-90]{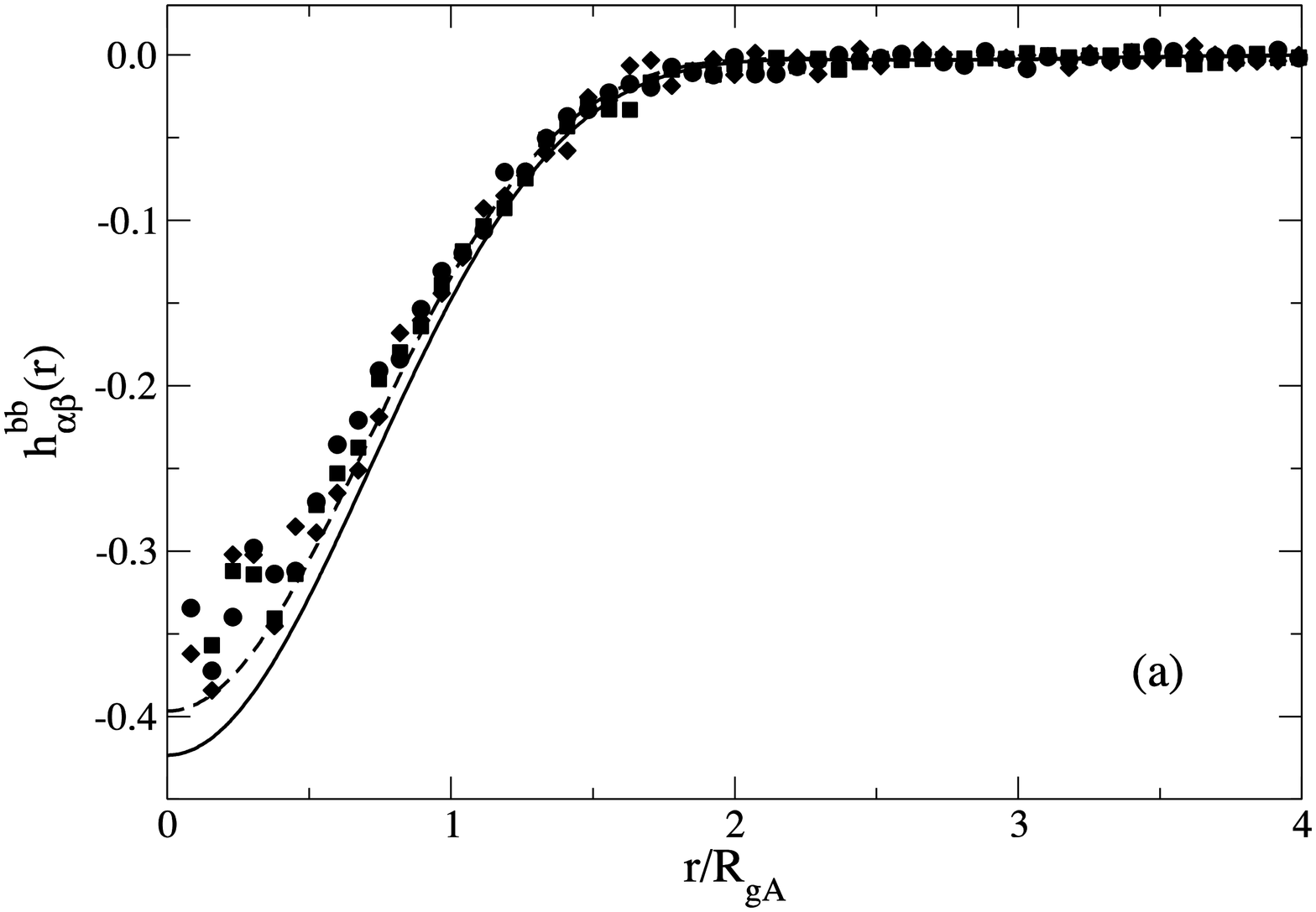}
\\
\includegraphics[scale=0.4,angle=-90]{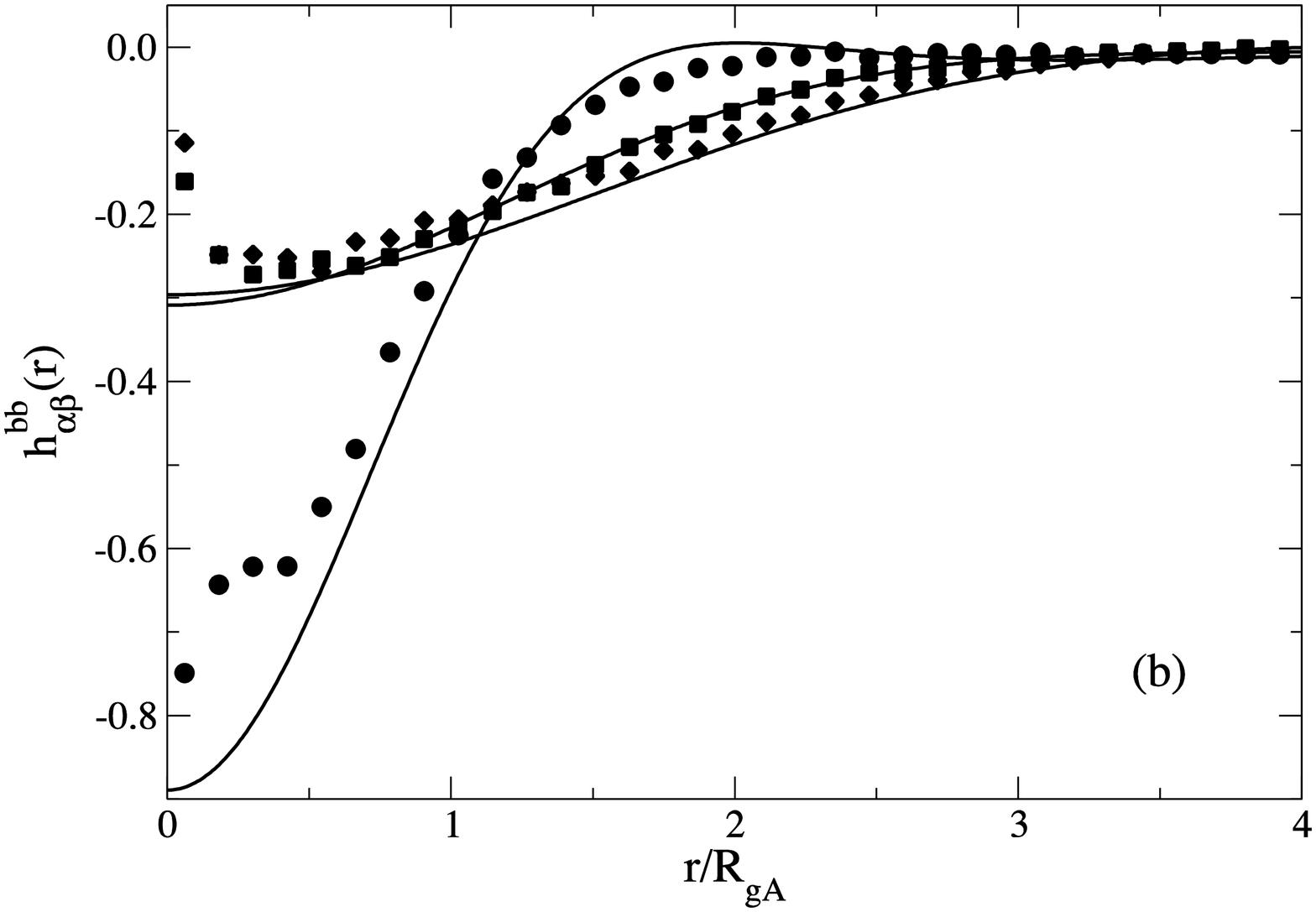}
\caption{}
\label{FG:HBBR}
\end{figure}

FIG.\ \ref{FG:HBBR}
Plot of $h^{bb}_{\alpha\beta}(r)$.  Data is arranged as in Fig.\
\ref{FG:HBBK}.\\

\begin{figure}
\centering
\includegraphics[scale=0.4,angle=-90]{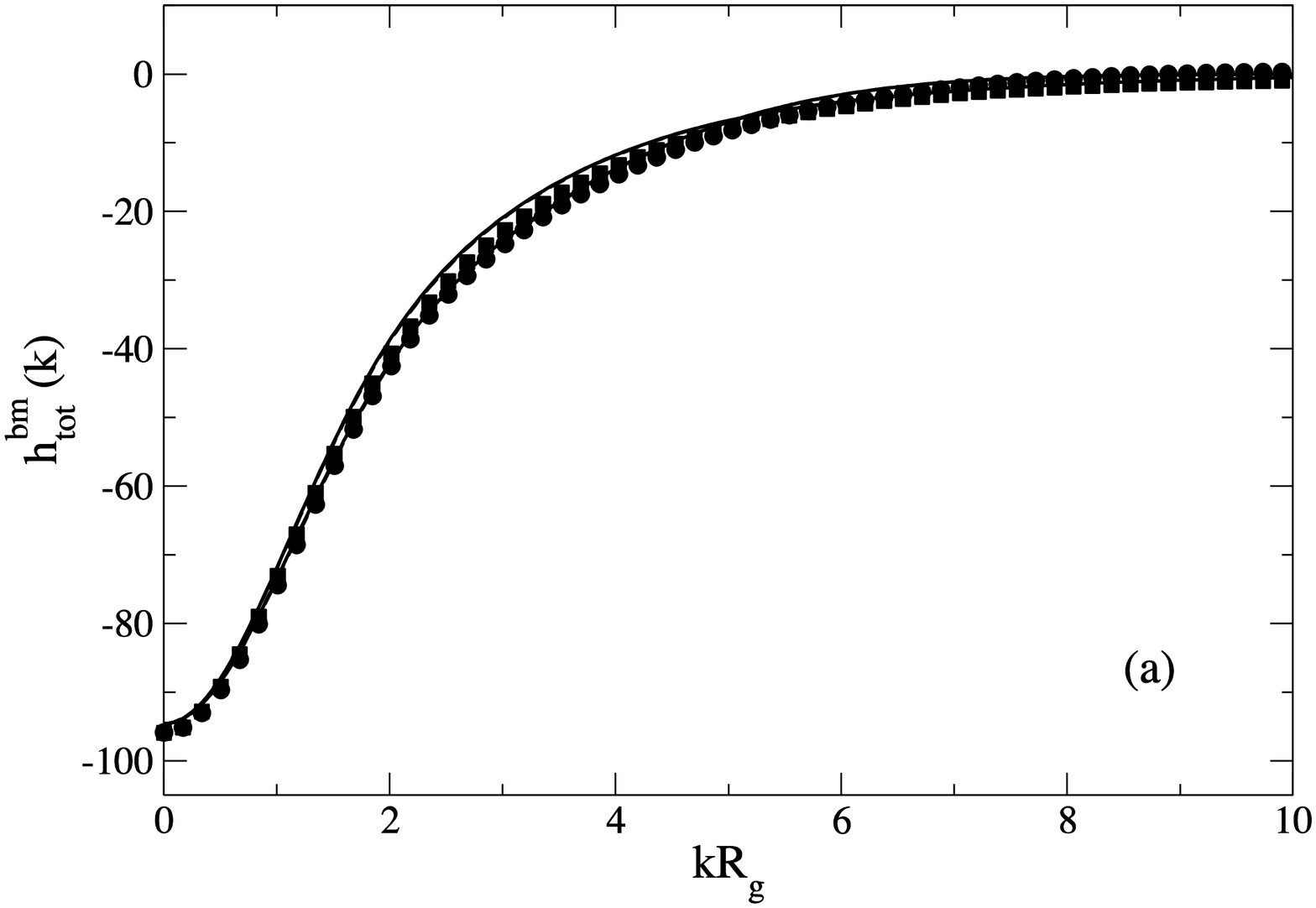}
\\
\includegraphics[scale=0.4,angle=-90]{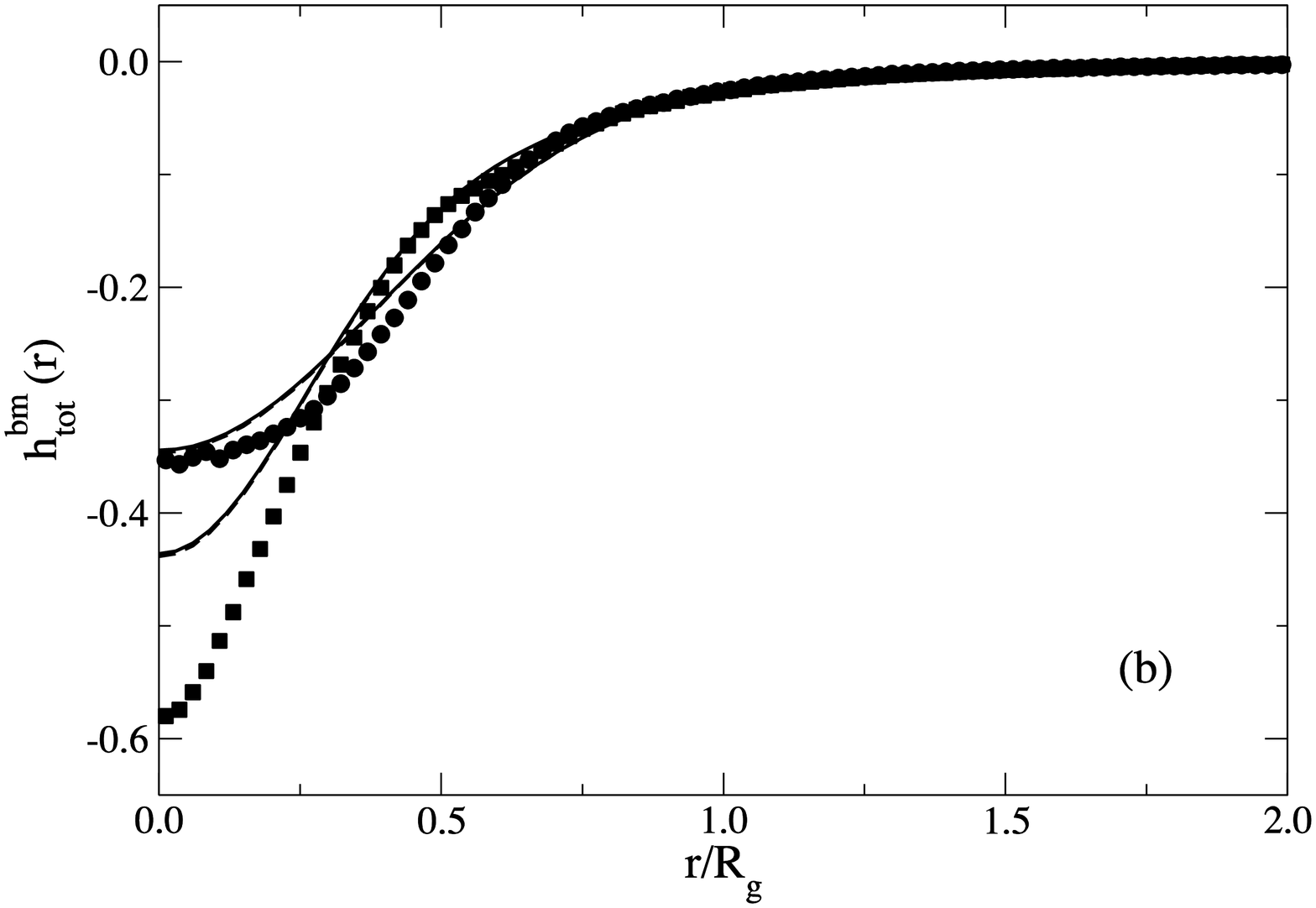}
\caption{}
\label{FG:HBMtt}
\end{figure}

FIG.\ \ref{FG:HBMtt}
Plot of $h^{bm}_{\mathit{tot}}(k)$ [Panel (a)] and
$h^{bm}_{\mathit{tot}}(r)$ [Panel (b)].  Lines are theoretical results
whereas symbols are data from united atom molecular dynamics
simulations: shown are the $f=0.50$ [circles] and $f=0.25$ [squares]
cases.  The dashed line is as in Fig.\ \ref{FG:HBMaK}.\\

\begin{figure}[]
\centering
\includegraphics[scale=0.4,angle=-90]{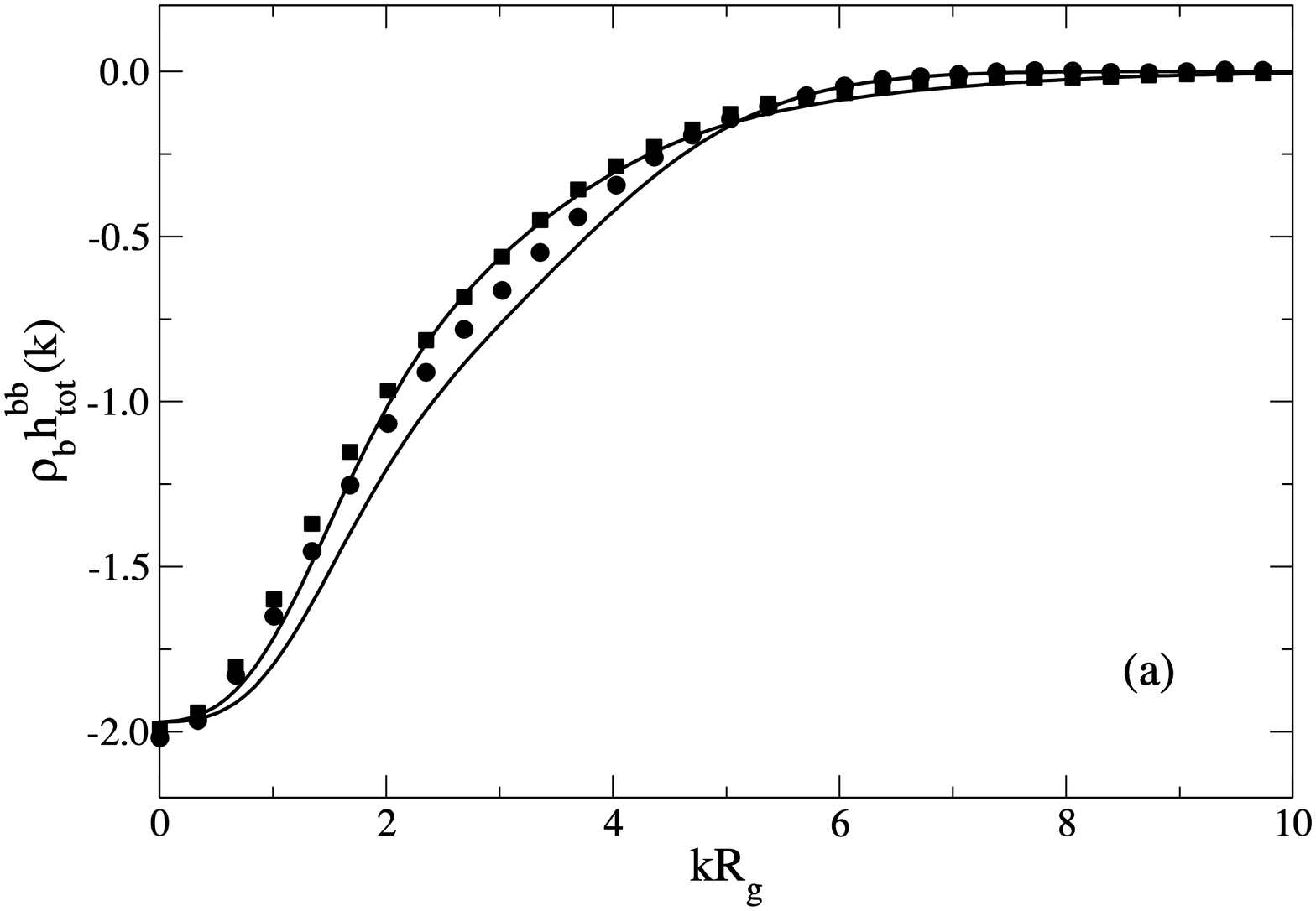}
\\
\includegraphics[scale=0.4,angle=-90]{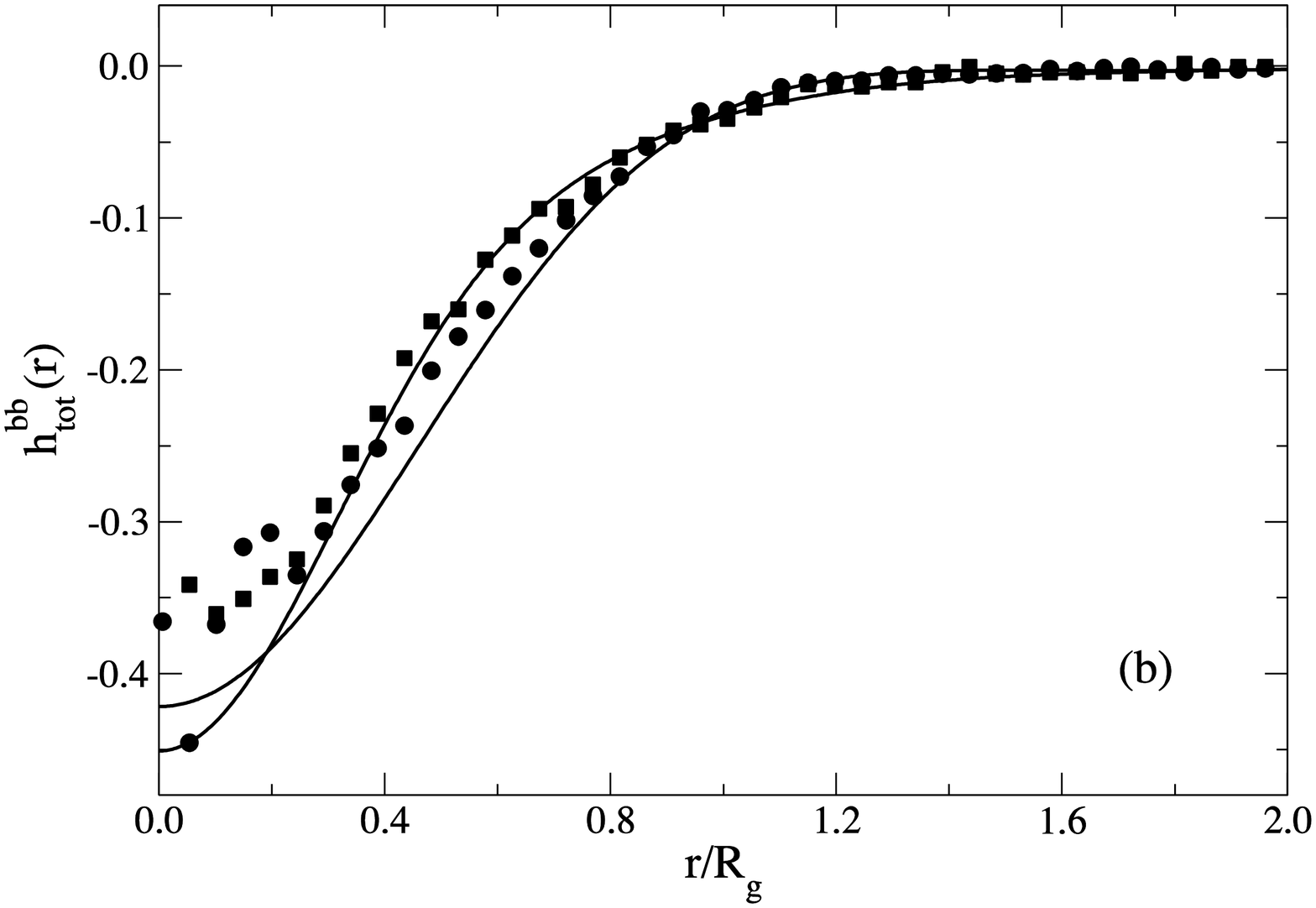}
\caption{}
\label{FG:HBBT}
\end{figure}

FIG.\ \ref{FG:HBBT}
Plot of $h^{bb}_{\mathit{tot}}(k)$ [Panel (a)] and
$h^{bb}_{\mathit{tot}}(r)$ [Panel (b)].  Data is arranged as in Fig.\
\ref{FG:HBMtt}.\\

\begin{figure}[]
\centering
\includegraphics[scale=0.4,angle=-90]{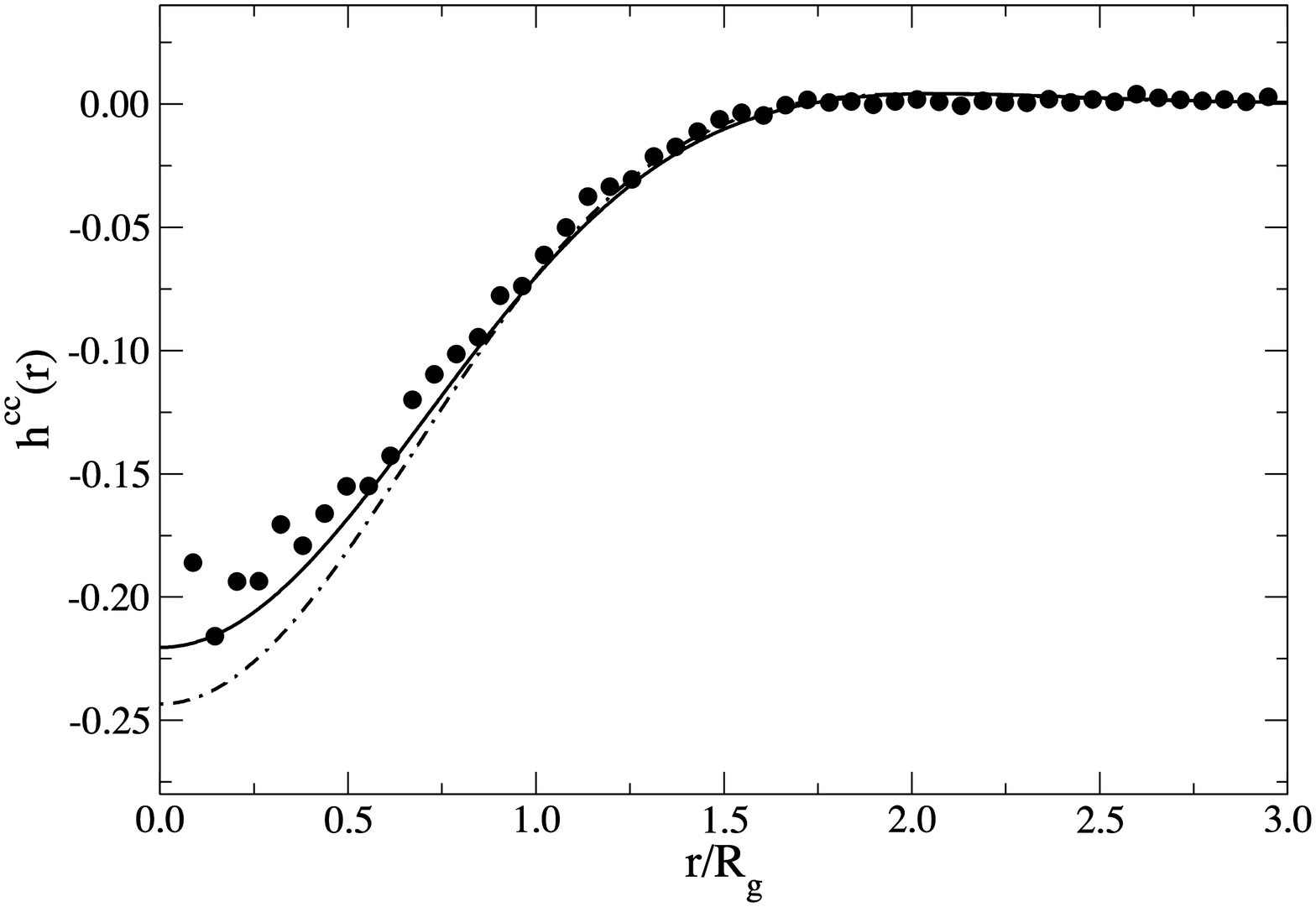}
\caption{}
\label{FG:HCCR}
\end{figure}

FIG.\ \ref{FG:HCCR}
Comparison of $h^{cc}(r)$ between theory and united atom simulation
data for athermal conditions.  Lines are theoretical results whereas
the symbols are data from united atom molecular dynamics simulations.
The representation obtained from the sum of
$\omega^{cm}_{\hphantom{\alpha}\alpha}(k)$ terms is also shown
[dot-dashed line].\\

\begin{figure}[]
\centering
\includegraphics[scale=0.4,angle=-90]{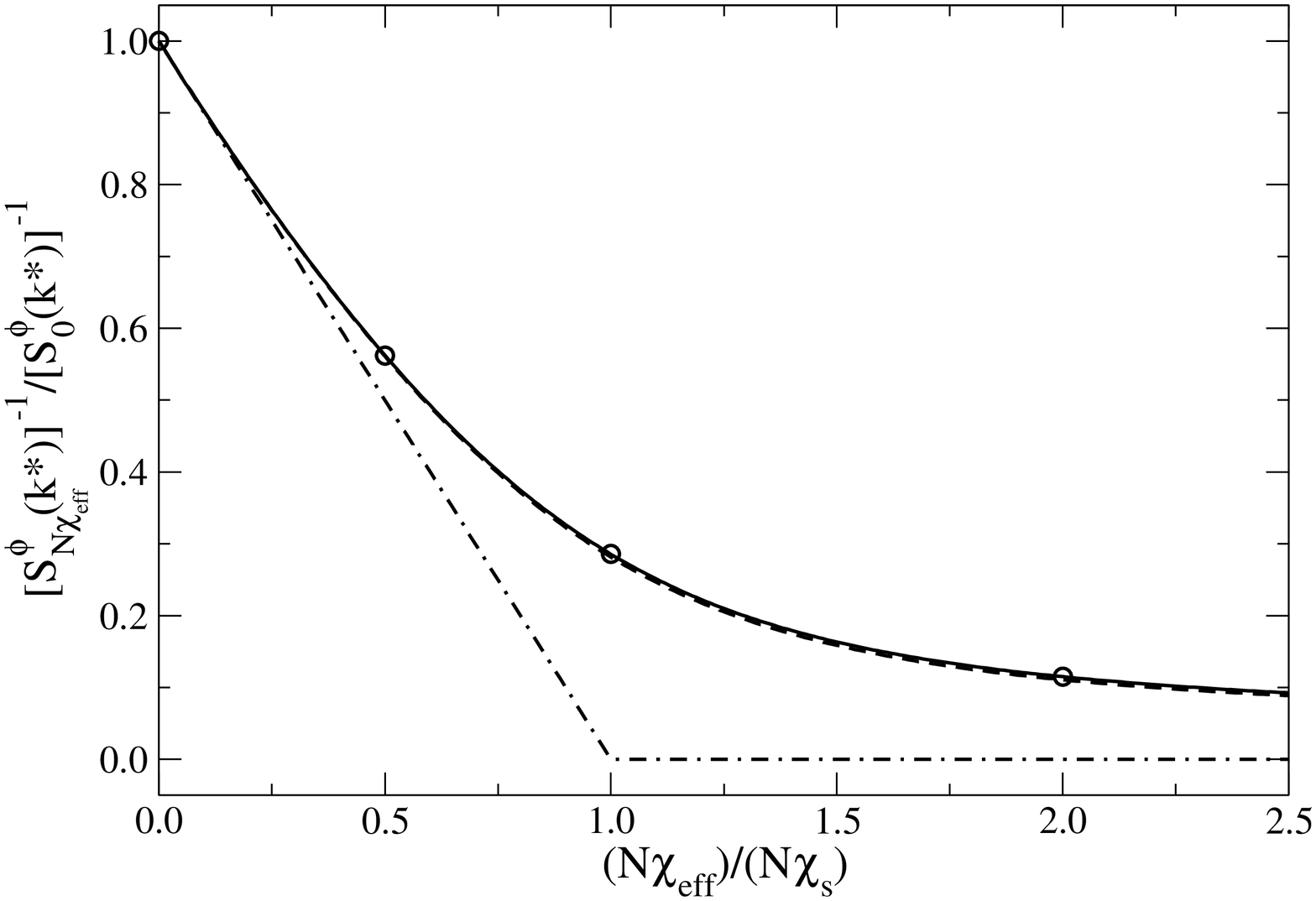}
\caption{}
\label{FG:CURV}
\end{figure}

FIG.\ \ref{FG:CURV}
Plot of the cooling curves calculated from the integral equation approach, for
a diblock copolymer system.  Shown are results for $f=0.50$
[solid line], $f=0.25$ [dashed line], and the mean-field behavior
[dot-dashed line].  The points sampled for the model calculations are
given by the circles.\\

\begin{figure}[]
\centering
\includegraphics[scale=0.4,angle=-90]{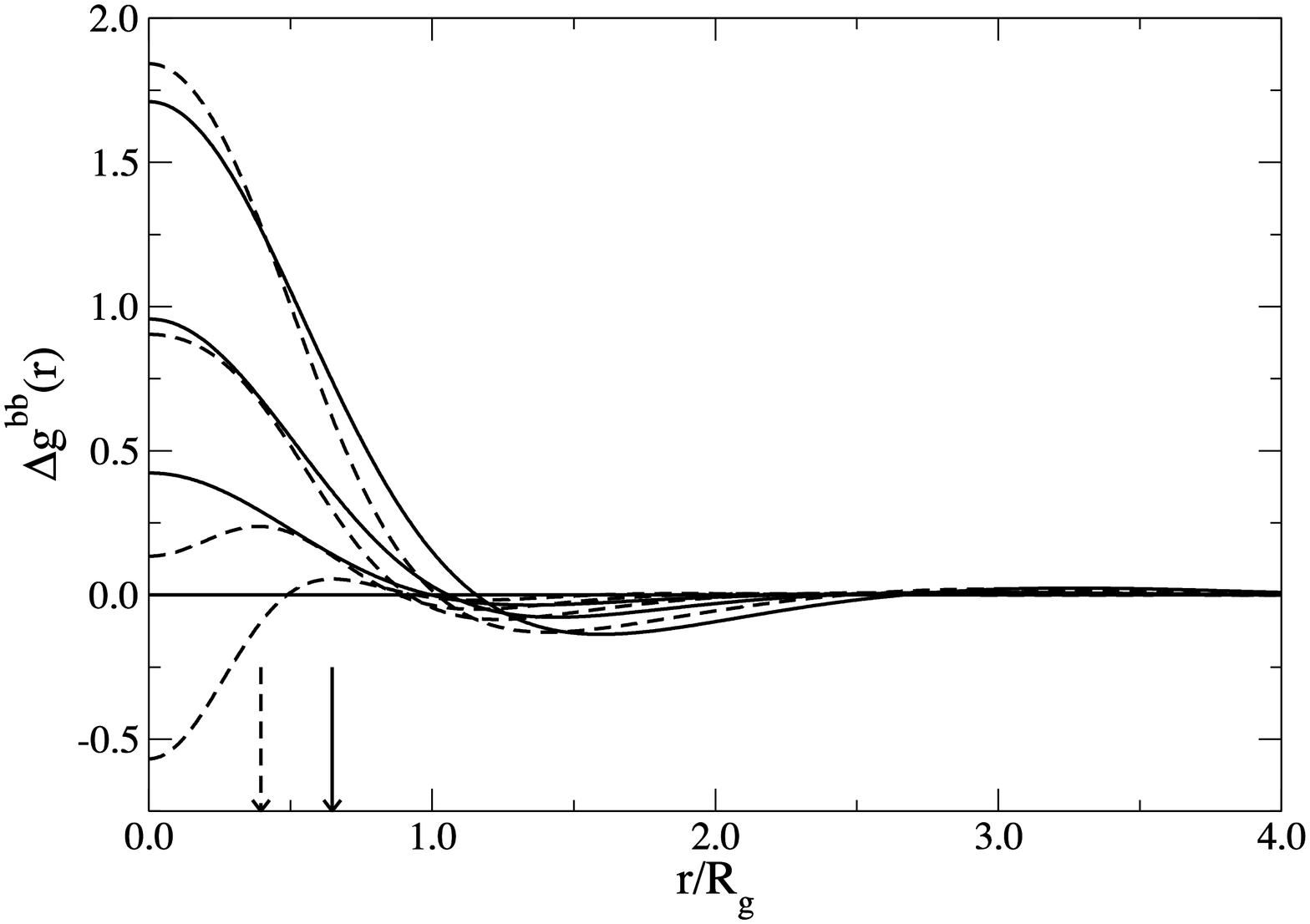}
\caption{}
\label{FG:HTMP}
\end{figure}

FIG.\ \ref{FG:HTMP}
Plot of $\Delta g^{bb}(r)$ as a function of the distance normalized by the 
polymer radius-of-gyration, for various temperatures. From bottom to
top: $N\chi_{\mathit{eff}}/N\chi_{hta,s} \in \{0.0, 0.5, 1.0, 2.0\}$.
Shown are the $f=0.50$ [solid lines] and $f=0.25$ [dashed lines]
cases.  The arrows indicate the respective size of $A$-blocks.\\

\newpage
% Figure 1

\newpage
% Figure 2

\newpage
% Figure 3

\newpage
% Figure 4

\newpage
% Figure 5

\newpage
% Figure 6

\newpage
% Figure 7

\newpage
% Figure 8

\newpage
% Figure 9

\newpage
% Figure 10
\end{document}